\documentclass[fleqn,usenatbib]{mnras}

\usepackage{newtxtext,newtxmath}

\usepackage[T1]{fontenc}

\DeclareRobustCommand{\VAN}[3]{#2}
\let\VANthebibliography\thebibliography
\def\thebibliography{\DeclareRobustCommand{\VAN}[3]{##3}\VANthebibliography}

\usepackage{graphicx}	
\usepackage{amsmath} 
 
\usepackage{amssymb}	

\usepackage{siunitx}
\usepackage{xcolor}
\usepackage{multirow}
\usepackage{multicol}
\usepackage{hyperref}

\DeclareSIUnit \parsec {pc}

\DeclareMathOperator\erf{erf}

\newcommand{\new}[1]{{#1}}
\newcommand{\neww}[1]{{#1}}

\newcommand{\fff}{$f_\text{ff}$}
\newcommand{\asy}{$\alpha_\text{sy}$}

\title[Ultraviolet-Radio Galaxy Fitting]{The Long and the Short of It: The Benefits and Leverage of Ultraviolet-Radio Galaxy Fitting}

\author[J. E. Thorne et al.]{
Jessica E. Thorne,$^{1}$\thanks{E-mail: jessica.thorne@icrar.org}
Aaron S. G. Robotham,$^{1}$
Sabine Bellstedt,$^{1}$
Luke J. M. Davies,$^{1}$
\\
$^{1}$ ICRAR, The University of Western Australia, 35 Stirling Highway, Crawley, WA 6009, Australia\\
}

\date{Accepted XXX. Received YYY; in original form ZZZ}

\pubyear{2022}

\begin{document}
\label{firstpage}
\pagerange{\pageref{firstpage}--\pageref{lastpage}}
\maketitle

\begin{abstract}
Traditionally, the far ultraviolet (FUV) to far-infrared (FIR) and radio spectral energy distributions (SEDs) of galaxies have been considered separately despite the common physical process shaping them. 
In this work, we explore the utility of simultaneously fitting FUV-radio SEDs using an extended version of the \textsc{ProSpect} SED fitting code considering contributions from both free-free and synchrotron emission.
We use a small sample of galaxies from the Deep Extragalactic VIsible Legacy Survey (DEVILS) and the Key Insights on Nearby Galaxies: a Far-Infrared Survey with \textit{Herschel} (KINGFISH) where high-quality and robust FUV-radio data are available to provide an ideal sample for testing a radio extension of \textsc{ProSpect}. 
As the parameterisation of the radio extension links the radio continuum to the FIR emission, we explore the benefit of using radio continuum measurements as a constraint on the energy balance between dust attenuation and emission. 
We find that for situations where MIR-FIR photometry is unavailable, including a 1.4\,GHz continuum measurement allows for improved accuracy in recovered star formation rates and dust luminosities of galaxies reducing the median uncertainty by 0.1 and 0.2\,dex respectively.
We also demonstrate that incorporating 3 and 10\,GHz measurements allows for further constraint on the energy balance and therefore the star formation rate and dust luminosity. 
This demonstrates the advantage of extending FUV-FIR SED fitting techniques to radio frequencies, especially as we move into an era where FIR surveys will remain limited and radio data become abundant (i.e. with the SKA and precursors). 
\end{abstract}





\section{Introduction}
Galaxy spectral energy distributions (SEDs) contain a wealth of information about the physical processes occurring within a galaxy and are shaped by contributions from stars, dust, gas, active galactic nuclei (AGN) etc \citep[see][]{WalcherFittingintegratedSpectral2011, ConroyModelingPanchromaticSpectral2013}. 
Each of these components contribute at different wavelengths, leaving their imprint on a galaxy SED.
As different emission mechanisms dominate in different wavelength regimes, certain wavelengths are used to measure certain astrophysical quantities of interest. 
For example, the stellar mass of galaxies is often measured in the optical and/or near infrared as emission from stars dominates at these wavelengths \citep{ZibettiResolvedstellarmass2009,TaylorGalaxyMassAssembly2011}. 
In addition, star formation rates (SFRs) can be measured at ultraviolet (UV) wavelengths as this regime is dominated by short-lived OB stars \citep{KennicuttStarFormationMilky2012}. 
However, this is made difficult due to the presence of dust which preferentially absorbs shorter wavelength photons (UV-optical) and re-emits at far infrared (FIR) wavelengths \citep{DraineInterstellarDustGrains2003}.
To account for the impact of dust, SFRs can also be measured using the combination of UV and infrared \citep{BellUnderstandingRapidDecline2005}, however, this measurement can be biased by dust heating from numerous low-mass stars.
As there is considerable interplay between galaxy components resulting in a significant impact on the SED, it has become more common to consider the components simultaneously \citep[e.g.][]{MAGPHYS,CIGALE,RobothamProSpectgeneratingspectral2020}. 

Many codes have been developed to measure key galaxy properties by self-consistently fitting the far ultraviolet (FUV) to FIR SEDs of galaxies.
Examples of these include \textsc{ProSpect} \citep{RobothamProSpectgeneratingspectral2020}, \textsc{magphys} \citep{MAGPHYS}, \textsc{cigale} \citep{CIGALE, BoquienCIGALEpythonCode2019}, \textsc{prospector} \citep{JohnsonStellarPopulationInference2021}, \textsc{beagle} \citep{ChevallardModellinginterpretingspectral2016}, and \textsc{bagpipes} \citep{CarnallInferringstarformation2018}.
Traditionally, SED fitting codes have been focused on measuring galaxy stellar masses and SFRs, however, advancements in photometric data quality have allowed for more reliable estimates of age, dust attenuation, and metallicity \citep{BellstedtGalaxyMassAssembly2020b,LowerHowWellCan2022,ThorneDEVILScosmicevolution2022}.

As with the FUV-FIR SEDs of galaxies, the shape of a galaxy's SED at radio frequencies is governed by the processes occurring within it including contributions related to star formation and the presence of an AGN. 
Observed radio emission can be used as a direct probe into the star-formation activity in `normal' galaxies (i.e. those without AGN), as radio luminosity is known to be directly proportional to recent star-formation rate \citep{CondonRadioemissionnormal1992,HaarsmaFaintRadioSources2000,BellOpticalNearInfraredProperties2003,HopkinsStarFormationRate2003,DaviesGalaxyMassAssembly2017,GurkanLOFARHATLASlowfrequency2018,SmithLOFARTwometreSky2021,HeesenNearbygalaxiesLOFAR2022}. 
In addition, the shape of radio SEDs, including the spectral indices \new{($\alpha$, where $S_\nu \propto \nu^{\alpha}$)}, can also be used to explore radiation laws and the cooling/heating mechanisms of the interstellar medium (ISM), and for some types of AGN can be used to probe the accretion activity and AGN timescales. 


Radio SEDs of star-forming galaxies (at 1.4\,GHz and higher frequencies) are generally the result of the combination of two distinct mechanisms, free-free emission and absorption, and synchrotron emission. 
These emission mechanisms arise from different physical processes in the galaxy.
As such, in order to robustly link the observed radio emission to a physical process, and therefore a star formation or AGN mechanism, they must first be deconstructed into their respective components. 

Synchrotron emission, the emission of photons from charged particles (usually electrons) spiraling around magnetic field lines at close to the speed of light, dominates the radio emission from most normal galaxies at frequencies below $\nu\sim$30\,GHz. 
Synchrotron emission in star-forming galaxies is thought to be produced by supernova remnants which accelerate electrons to relativistic speeds \citep{HarwitInfraredgalaxiesEvolutionary1975}. 
The supernova rate of galaxies is inherently linked to the birth of high-mass stars, as only stars more massive than $\sim8M_\odot$ can produce the Type II and Type Ib core-collapse supernovae whose remnants are thought to accelerate most of the electrons in normal galaxies \citep{CondonRadioemissionnormal1992}. 
This means that the non-thermal radio luminosity provides a measure of the current star formation on timescales of $\sim100$\,Myr for an assumed supernova rate and IMF \citep{CondonRadioSourcesStar2002}.

Thermal free-free emission is powered by the ionisation of H\,{\sc ii} regions by UV flux from the same high-mass stars that eventually produce the supernovae remnants associated with non-thermal synchrotron emission (but at an earlier stage in their lifecycle). 
This means that, unlike synchrotron, free-free emission is a direct, near-instantaneous tracer of star formation. 
Despite this, free-free emission is rarely used as a radio-continuum SFR indicator  as 
at the low frequencies typically targeted by large radio surveys, the radio continuum is overwhelmingly dominated by synchrotron emission. 
Isolating the free-free contribution requires either model fitting using well-sampled radio-continuum SEDs \citep{PriceNewResultsRadio1992,TabatabaeiRadioSpectralEnergy2017} or high-frequency observations ($\nu > 20$\,GHz) where the contribution from synchrotron emission is much lower \citep{MurphyStarFormationRadio2012}. 

One of the challenges with studying the SEDs of galaxies at radio frequencies is that most surveys target a single radio frequency or band (mostly 1.4GHz) and differences in sensitivity, resolution, and data processing techniques prohibit coherent radio SED analysis of many galaxies \citep{TabatabaeiRadioSpectralEnergy2017}. 
In the coming decades, the Square Kilometer Array (SKA) and its pathfinder projects will significantly increase the amount of radio data available for galaxies with robust multiwavelength imaging at improved resolution and sensitivity. 
However, many of the SKA and pathfinder surveys \citep{MeyerExploringHIUniverse2009,NorrisEMUEvolutionaryMap2011,JarvisMeerKATInternationalGHz2016,KoribalskiWALLABYSKAPathfinder2020,McConnellRapidASKAPContinuum2020,ShimwellLOFARTwometreSky2022,deGasperinLOFARLBASky2023} will still be focused on a single radio band and multiple surveys will have to be combined to study radio SEDs at multiple frequencies.

To make the most of this wealth of new radio data, unpacking the relative contributions of free-free and synchrotron emission in different galaxy types will be vital.
This cannot be done using one or two data points alone and ideally requires measurements at as many different radio frequencies as possible. 

An alternative to requiring many different radio measurements is to simultaneously consider the FUV-FIR and radio regimes. 
As discussed previously, radio emission, both free-free and synchrotron, is inherently linked to star formation activity as is the shape of galaxy SEDs at FUV-NIR and FIR wavelengths through both the slope of the FUV-NIR regime and the re-emission by dust at FIR wavelengths. 
This means that the FUV-FIR SEDs of galaxies can be used to help constrain the normalisation of the free-free and synchrotron radio continuum and aid in their separation. 

Although it is theoretically straightforward to couple the observed radio emission to the star-formation rate or FIR luminosity derived from SED fitting, this has only previously been implemented in the SED-fitting codes \textsc{cigale} \citep{CIGALE,BoquienCIGALEpythonCode2019} \new{and \textsc{magphys} \citep{daCunhaALMASurveySubmillimeter2015}}. 
However, the \textsc{cigale} implementation \citep{DeyLowfrequencyRadioContinuum2022} uses only a single power-law component to model the radio continuum despite the known differences in spectral indices between the free-free and synchrotron components. 
\citet{Drouartnaturelikelyredshift2021} and \citet{SeymourHSTWFC3Grism2022} also simultaneously fit the FUV-radio SED of a high redshift radio galaxy using a bespoke tool to fit \textsc{pégase} templates \citep{FiocPEGASEcodemodeling2019} for the FUV-FIR in combination with an analytical function at radio frequencies.
As with the \textsc{cigale} implementation, the analytic function used by \citet{Drouartnaturelikelyredshift2021} and \citet{SeymourHSTWFC3Grism2022} does not separate the contributions from free-free and synchrotron emission. 
Using a sample of low-redshift galaxies, \cite{TabatabaeiRadioSpectralEnergy2017} demonstrated that spectral indices derived using a single power-law component can differ significantly from the synchrotron spectral indices derived using a two-component model. 
Hence, the \textsc{cigale} and \citet{Drouartnaturelikelyredshift2021} implementations are limited in their ability to provide a robust physical interpretation of the observed radio emission.
The assumption of a single power-law component also prevents the estimation of the relative contributions from the free-free and synchrotron components.
\new{The radio extension of \textsc{magphys} described in \cite{daCunhaALMASurveySubmillimeter2015} separates the radio emission into free-free and synchrotron contributions but uses fixed spectral indices for both components and a fixed free-free contribution of 10 per cent at 20cm. 
However, this implementation is applied to a sample of galaxies with a single radio measurement at 1.4\,GHz and therefore a sophisticated decomposition of the radio spectrum would be unnecessary. 
}

In this work, we extend the \textsc{ProSpect} FUV-FIR SED-fitting code to radio frequencies considering both the contribution from free-free and synchrotron processes. 
We apply the extended \textsc{ProSpect} code to a small sample of galaxies to demonstrate the utility in simultaneously fitting FUV-radio SEDs and highlight its viability to be applied to much larger samples that will be available in the SKA era. 
this work is structured as follows. 
Section~\ref{sec:RadioData} describes the additional data sets used to supplement the DEVILS data. 
We describe the radio extension to \textsc{ProSpect} and its implementation in Section~\ref{sec:RadioSED}. 
The resulting radio SED parameters and their correlation with other galaxy properties are explored in Section~\ref{sec:RadioSEDParameters} and the 1.4\,GHz Luminosity -- SFR relation for the sample is presented in Section~\ref{sec:L14SFR}.
In Section~\ref{sec:energybalance} we explore the use of radio continuum data in providing constraint on dust emission in instances where FIR data is not available. 
We summarise our results in Section~\ref{sec:radiosummary}.
Throughout this work, we use a \cite{ChabrierGalacticStellarSubstellar2003} IMF and all magnitudes are quoted in the AB system. 
We adopt the \cite{PlanckCollaborationPlanck2015results2016} cosmology with $H_0 = 67.8 \, \si{\kilo \meter \per \second \per \mega \parsec}$, $\Omega_{M} = 0.308$ and $\Omega_\Lambda = 0.692$. 

\section{Data}\label{sec:RadioData}

\subsection{Deep Extragalactic VIsible Legacy Survey}
For this work, we use the Deep Extragalactic VIsible Legacy Survey (DEVILS; \citealt{DaviesDeepExtragalacticVIsible2018}). 
DEVILS is an optical spectroscopic redshift survey using the Anglo-Australian Telescope specifically designed to have high spectroscopic completeness over a large redshift range ($z < 1$) in three well-studied extragalactic fields: XMM-LSS/D02, ECDFS/D03, and COSMOS/D10 covering a total of 4.5 deg$^2$. 
As per \cite{ThorneDeepExtragalacticVIsible2021, ThorneDeepExtragalacticVIsible2022,ThorneDEVILScosmicevolution2022}, we use the spectroscopic and photometric data from the D10-COSMOS field as it is the deepest field. 
We use the DEVILS photometry catalogue derived using the \textsc{ProFound} source extraction code \citep{RobothamProFoundSourceExtraction2018} and described in depth by \cite{DaviesDeepExtragalacticVIsible2021}.
\textsc{ProFound} is used for source finding and photometry extraction consistently across 22 bands spanning the FUV-FIR (1500\,\AA-500\,$\mu$m) and includes GALEX \textit{FUV NUV} \citep{ZamojskiDeepGALEXImaging2007}, CFHT \textit{u} \citep{CapakFirstReleaseCOSMOS2007},  Subaru HSC \textit{griz} \citep{AiharaSeconddatarelease2019},  VISTA \textit{YJHK$_{s}$} \citep{McCrackenUltraVISTAnewultradeep2012}, Spitzer \textit{IRAC1 IRAC2 IRAC3 IRAC4 MIPS24 MIPS70} \citep{LaigleCOSMOS2015CATALOGEXPLORING2016,SandersSCOSMOSSpitzerLegacy2007}, and Herschel \textit{P100 P160 S250 S350 S500} \citep{LutzPACSEvolutionaryProbe2011,OliverHerschelMultitieredExtragalactic2012} bands.
The DEVILS redshift catalogues spanning $0<z<8$ have been compiled using photometric, grism, and spectroscopic redshifts and are described in \cite{ThorneDeepExtragalacticVIsible2021}.

To extend the FUV-FIR DEVILS photometry to longer wavelengths we use data from the MeerKAT International Gigahertz Tiered Extragalactic Explorations (MIGHTEE; \citealt{JarvisMeerKATInternationalGHz2016}) and COSMOS-XS \citep{AlgeraMultiwavelengthAnalysisFaint2020,vanderVlugtUltradeepMultibandVLA2021} surveys to select a sample of galaxies with high quality FUV-radio data. 
The COSMOS field is also covered by larger area surveys such as the VLA-COSMOS 3\,GHz Large Project \citep{SmolcicVLACOSMOSGHzLarge2017} and the VLA-COSMOS 1.4\,GHz Deep Project \citep{SchinnererVLACOSMOSSurveyII2007} however the 3\,GHz survey has significantly finer resolution (0.75 arcseconds) than the other surveys and is known to resolve out flux for extended sources.
Additionally, the COSMOS 1.4\,GHz Deep Project has been superseded by MIGHTEE. 

\subsubsection{MeerKAT International Gigahertz Tiered Extragalactic Explorations (MIGHTEE)}\label{sec:MIGHTEE}
The MIGHTEE large survey project will survey four well-studied deep extragalactic fields, totaling 20 deg$^2$ to $\mu$Jy sensitivity at GHz frequencies using the MeerKAT telescope \citep{JarvisMeerKATInternationalGHz2016}. 
MIGHTEE uses simultaneous continuum \citep{HeywoodMIGHTEETotalintensity2022}, polarimetry (Sekhar et al. in prep.), and spectral line \citep{MaddoxMIGHTEEHIemissionproject2021} measurements with MeerKAT's L-band (870-1670 MHz) receivers. 

In this work, we make use of the Level-1 `Early Science' continuum catalogue for the COSMOS field presented in \cite{HeywoodMIGHTEETotalintensity2022} \new{centered at 1.284\,GHz.}
This consists of a single pointing in the COSMOS field, observed for 25 hours for an on-source time of 17.45 hours. 
MIGHTEE continuum data are imaged twice, with a Briggs’ robust parameter of 0.0 and -1.2, resulting in a higher sensitivity image as well as a finer angular resolution image.
The COSMOS pointing reaches a thermal noise (measured away from the main lobe of the primary beam) of 1.9\,$\mu$Jy beam$^{-1}$ in the robust 0.0 image, with an angular resolution of 8.6 arcsec. 
The robust -1.2 image reaches 6\,$\mu$Jy beam$^{-1}$ with an angular resolution of 5 arcsec.
Source finding was completed using \textsc{pybdsf} \citep{MohanPyBDSFPythonBlob2015} to locate and characterise components. 

We also make use of the source classifications from \cite{WhittamMIGHTEEnatureradioloud2022} to identify radio-loud AGN in the MIGHTEE sample. 
RLAGN were selected using the infrared--radio correlation (IRRC) to identify sources with significantly more radio emission than would be expected from star-formation alone. 
The IRRC can be quantified by the parameter $q_\text{IR}$ which is defined as follows:
\begin{equation}
    q_\text{IR} = \log_{10} \frac{L_\text{IR} (W) / 3.75\times10^{12} \text{Hz}}{L_{1.4\text{\,GHz}} (\text{W Hz}^{-1} )},
\end{equation}
where $L_\text{IR}$ is the total infrared luminosity between 8-1000\,$\mu$m. 
This is divided by the central frequency of $3.75\times10^{12}$\,Hz ($80\,\mu$m) so that $q_\text{IR}$ is a dimensionless quantity. 
They use the stellar mass and redshift dependent IRRC from \citet{Delvecchioinfraredradiocorrelationstarforming2021} and select sources which lie more than 0.43\,dex below the best fit correlation as having a radio excess (this corresponds to 2$\,\sigma$, where $\sigma$ is the intrinsic scatter in the relation). 
301 MIGHTEE sources were not classified as radio loud as they did not have reliable constraints on the total infrared luminosity, however, none of these objects are in our final sample. 

\subsubsection{COSMOS-XS}\label{sec:COSMOSXS}
The COSMOS-XS survey \citep{AlgeraMultiwavelengthAnalysisFaint2020,vanderVlugtUltradeepMultibandVLA2021} combines two single VLA pointings at sub-microJansky depth in the COSMOS field at X-band (10\,GHz, 90 hours) and S-band (3\,GHz, 100 hours). 
The X band covers a bandwidth of 4096\,MHz centered at 10\,GHz while the S band covers a bandwidth of 2048\,MHz centered at 3\,GHz.

Imaging of both data sets was performed using the standalone imager \textsc{WSCLEAN} \citep{OffringaWSCLEANimplementationfast2014}, incorporating \textit{w}-stacking to account for the non-coplanarity of the baselines. 
Both images were created via Briggs weighting, with a robust parameter of 0.5.
The images at both frequencies have a resolution of $\sim2''$, which is large enough to avoid resolving out faint sources. 
The area overlap between the two frequency images covers approximately 30 arcmin$^2$, all of which lies within the DEVILS D10 field and MIGHTEE coverage.

\subsubsection{Sample Selection}
To generate a combined catalogue spanning the FUV-radio regimes, we position match the MIGHTEE catalogue to our catalogue with a 2'' radius using the \texttt{coordmatch} function from the \textsc{R} \textsc{celestial} package\footnote{\url{https://github.com/asgr/celestial}}and find radio counterparts for 6,845 of our sources. 
We visually inspect the optical and radio source positions to ensure that we are matching the radio object to the correct optical object.
We then position match the COSMOS-XS sources to this combined catalogue again using a 2'' radius.
This results in a sample of 52 DEVILS galaxies with measurements at 1.4, 3, and 10\,GHz. 

To ensure all sources have secure redshifts we limit our sample to only objects with spectroscopic redshifts in the DEVILS-D10 catalogue. 
As the radio extension implemented in \textsc{ProSpect} (described below) uses FIR emission to predict the free-free emission at 1.4\,GHz we also want to ensure that our FIR emission is well constrained. 
To do this we limit our sample to only objects with a flux density measurement above \new{$3.63 \times 10^{-5}$\,Jy (20th mag)} in at least one FIR band. 

These implemented cuts result in a sample of 33 galaxies with spectroscopic redshifts and high-quality FUV-radio photometry. 

\begin{table*}
    \centering
    \caption[The parameters used by \textsc{ProSpect} in this work.]{The parameters used by \textsc{ProSpect} in this work. We list the parameter name, a brief description, whether it is fit in linear or logarithmic (log) space or if it is fixed, the range of allowed values and any imposed non-uniform prior.}
    \label{tab:ParametersRadio}
    \begin{tabular}{l p{6cm} l l l l}
    \hline
    Parameter & Description & Type & Units & Values & Prior \\
    \hline
    \texttt{mSFR} &  \parbox{3.2cm}{peak star formation rate} &  log & $M_\odot\,\text{yr}^{-1}$ & [-3,4] & \\
    \texttt{mpeak} &  lookback time when peak star formation occurred & linear & Gyr & [-2,13.38] & \\
    \texttt{mperiod} &  width of the SFH & log &  Gyr & [$\log_{10}(0.3)$,2] &  $100 \erf (\texttt{mperiod}+2) - 100$ \\
    \texttt{mskew} &  skewness of the SFH & linear  & & [-0.5,1] & \\
    \texttt{Zfinal} &  final gas-phase metallicity & log & & [-4, -1.3] &\\  
    \hline
     \texttt{alpha\_SF\_birth}&  Power law of the radiation \newline field heating birth cloud dust  & linear & & [0,4] & $\exp{(-\frac{1}{2} (\frac{\alpha_\text{birth} + 2}{1})^2 ) }$ \\
    \texttt{alpha\_SF\_screen}&  Power law of the radiation \newline field heating general ISM dust  & linear & & [0,4] & $\exp{(-\frac{1}{2} (\frac{\alpha_\text{screen} + 2}{1})^2 ) }$ \\
       \texttt{tau\_birth} & optical depth of the birth clouds & log & & [-2.5,1] & $\exp{(-\frac{1}{2} (\frac{\tau_\text{birth} - 0.2}{0.5})^2 ) }$ \\
    \texttt{tau\_screen} &  optical depth of the general ISM & log & & [-5,1]  & $-20\erf(\tau_\text{screen}-2) $  \\
    \hline
     \texttt{AGNan} & angle of observation & linear & deg & [0.001,89.990] & \\
         \texttt{AGNlum} & bolometric luminosity of AGN source & log & erg s$^{-1}$ & [35,49] & \\
           \texttt{AGNta} & optical depth tau & log & & [-1,1] & \\
        \texttt{AGNrm} &  outer to inner torus radius ratio & fixed & & 60 &\\
         \texttt{AGNbe} & beta dust parameter & fixed & & -0.5 &  \\
         \texttt{AGNal} & gamma dust parameter & fixed  & & 4.0 & \\
         \texttt{AGNct} & opening angle of torus & fixed & deg & 100 & \\
         \hline
    \texttt{ff\_frac\_SF} & fraction of free-free radio emission contribution at 1.4 GHz& log &  &[-5,-0.5] & \\
    \texttt{sy\_power\_SF} & \new{spectral index} of the synchrotron radio emission  & linear &  & [-2.2,0] & \\
    \texttt{ff\_power\_SF} & \new{spectral index} of the free-free radio emission & fixed & &  -0.1 & \\
    \hline
    \end{tabular}
\end{table*}

\subsection{KINGFISHER}
To supplement our higher redshift DEVILS galaxies with systems in the local Universe, we use the union of the KINGFISH (Key Insights on Nearby Galaxies: A Far-Infrared Survey with \textit{Herschel}; \citealt{KennicuttKINGFISHKeyInsights2011,DaleHerschelFarinfraredSubmillimeter2012}) and SINGS (\textit{Spitzer} Infrared Nearby Galaxies Survey, \citealt{KennicuttSINGSSIRTFNearby2003,DaleInfraredSpectralEnergy2005, DaleUltraviolettoRadioBroadbandSpectral2007}) samples. 
The galaxies in the KINGFISH sample were selected to cover a wide range of galaxy properties and ISM conditions found in the nearby Universe to better understand the physical processes linking star formation and the ISM. 

SINGS recovered infrared imaging and spectroscopy for 75 nearby galaxies spanning a broad range of galaxy properties and star formation environments, while KINGFISH provided far-infrared/submillimeter data for a sample of 61 nearby galaxies, of which 57 are also SINGS targets. 
Recently, \cite{DaleUpdated34bandPhotometry2017} published updated global photometry for the 79 galaxies that comprise the union of the KINGFISH and SINGS samples, with photometry spanning the FUV to submillimeter. 
We use all available FUV-NIR photometry, namely \textit{GALEX} FUV \& NUV \citep{GildePazGALEXUltravioletAtlas2007}, Harris \textit{B,V,R,I} \citep{DaleUltraviolettoRadioBroadbandSpectral2007} , SDSS \textit{u,g,r,i,z} \citep{AlamEleventhTwelfthData2015}, 2MASS \textit{J,H,Ks} \citep{Jarrett2MASSLargeGalaxy2003}, \textit{Spitzer} IRAC Channels 1-4, WISE bands 1-4 \citep{WrightWidefieldInfraredSurvey2010,JarrettExtendingNearbyGalaxy2013}, \textit{Spitzer} MIPS 24,70 and 160\,$\mu$m, \textit{Herschel} PACS 70,100, and 160\,$\mu$m, \textit{Herschel} SPIRE 250, 350, and 500\,$\mu$m \citep{DaleHerschelFarinfraredSubmillimeter2012} and include 850\,$\mu$m measurements from James Clerk Maxwell Telescope (JCMT, \citealt{HollandSCUBAcommonusersubmillimetre1999}) SCUBA or \textit{Planck} High-Frequency Instrument (HIFI, \citealt{PlanckCollaborationPlanck2015results2016}) where available.
The photometry provided by \citet{DaleUpdated34bandPhotometry2017} do not include corrections for galactic dust extinction, so we correct for this using the $E(B-V)$ values provided and the R(V) dependent curve presented in \cite{FitzpatrickCorrectingEffectsInterstellar1999}\footnote{This is done using the \textsc{dust\_extinction} python package (\url{https://github.com/karllark/dust_extinction}}. 

To extend these FUV-FIR SEDs to radio frequencies we use the measurements presented in \citet{TabatabaeiRadioSpectralEnergy2017} spanning $1.36-10$\,GHz. 
From the 61 galaxies in the KINGFISH sample, 50 galaxies with declinations $\geq -21\deg$ were selected and formed the KINGFISHER (KINGFISH galaxies Emitting in Radio) sample. 
For this sample, most radio measurements were made using the Effelsberg 100-m radio telescope, however, these data are supplemented with measurements using the Very Large Array (VLA), Westerbork Synthesis Radio Telescope, and Green Bank Telescope (see \citealt{TabatabaeiRadioSpectralEnergy2017} for more details).
Most galaxies have measurements at 10.7, 4.8, and 1.36-1.4\,GHz however some galaxies have not been observed at all three frequencies. 
Additional data has been included at 8.4, 5, and 2.7\,GHz where available.  
The flux densities provided by \citet{TabatabaeiRadioSpectralEnergy2017} were integrated up to the optical radius in order to be consistent with measurements in the IR.

\subsubsection{Sample Selection}
To ensure good constraint on the normalisation and shape of the SED at radio frequencies we remove upper limits and restrict our analysis of the KINGFISHER sample to just galaxies with at least three radio data points. 
To ensure that we are using the \new{correct bandpass} we also do not use data points where the telescope or band used for a measurement is not explicitly stated in the sample catalogues or papers. 
Finally, we also remove galaxies with an $E(B-V) >0.1$ to ensure that the galactic dust extinction corrections do not bias our results. 
These cuts result in a sample of 34 galaxies.

\section{Methods}\label{sec:RadioSED}
To fit the FUV-radio SEDs, we extend the SED fitting technique described in \citet{ThorneDeepExtragalacticVIsible2022} to also include a flexible radio extension. 
We briefly outline the \textsc{ProSpect} implementation from \citet{ThorneDeepExtragalacticVIsible2022} and then describe the extension to radio frequencies. 

\subsection{\textsc{ProSpect}}
We use the \textsc{ProSpect} SED fitting code \citep{RobothamProSpectgeneratingspectral2020}, with the \cite{BruzualStellarpopulationsynthesis2003} stellar templates, \cite{ChabrierGalacticStellarSubstellar2003} IMF and the \cite{CharlotSimpleModelAbsorption2000} dust attenuation and \cite{DaleTwoParameterModelInfrared2014} dust re-emission models. 
In our analysis, we use the \texttt{massfunc\_snorm\_trunc} parameterisation for the star formation history, which takes the form of a skewed Normal distribution, with the peak position (\texttt{mpeak}), peak SFR (\texttt{mSFR}), SFH width (\texttt{mperiod}), and SFH skewness (\texttt{mskew}) set as free parameters. 
The SFH is anchored to 0 at a lookback time of 13.4 Gyr, selected to be the age at which galaxies start forming (equivalent to $z=11$, \citealt{OeschREMARKABLYLUMINOUSGALAXY2016}).

In addition to the five free parameters specifying the star formation and metallicity histories, we include four free parameters to describe the contribution of dust to the SED. 
Within \textsc{ProSpect} the dust is assumed to exist in two forms; in birth clouds formed around young stars (age$<10^{7}\,$yr), or distributed as a screen in the ISM. 
For each of these components, we include two free parameters, describing the dust opacity (\texttt{tau\_screen}, \texttt{tau\_birth}), and the dust radiation field intensity (\texttt{alpha\_screen}, \texttt{alpha\_birth}). 
Figure 3 of \cite{ThorneDeepExtragalacticVIsible2021} shows the impact of each parameter on a generated galaxy SED.

We also include an AGN component by incorporating the model outlined in \cite{FritzRevisitinginfraredspectra2006} and \cite{FeltreSmoothclumpydust2012}. 
This models the primary source as a composition of power-laws, with different spectral indices as a function of the wavelength. 
To model the contribution from the torus, the \cite{FritzRevisitinginfraredspectra2006} model uses a simple but realistic torus geometry, a flared disc, and a dust grain distribution function including a full range of grain sizes and assumes that the dust in the AGN torus is smoothly distributed. 
Within \textsc{ProSpect} we model the AGN contribution by fitting the luminosity of the central source (\texttt{AGNlum}), optical depth at 9.7$\mu$m (\texttt{AGNta}), and angle of observation (\texttt{AGNan}). 
We also re-attenuate the emission from the central source and dust torus through the general ISM screen (see figure 1 of \citealt{RobothamProSpectgeneratingspectral2020}).

\begin{figure*}
    \centering
    \includegraphics[width = \linewidth]{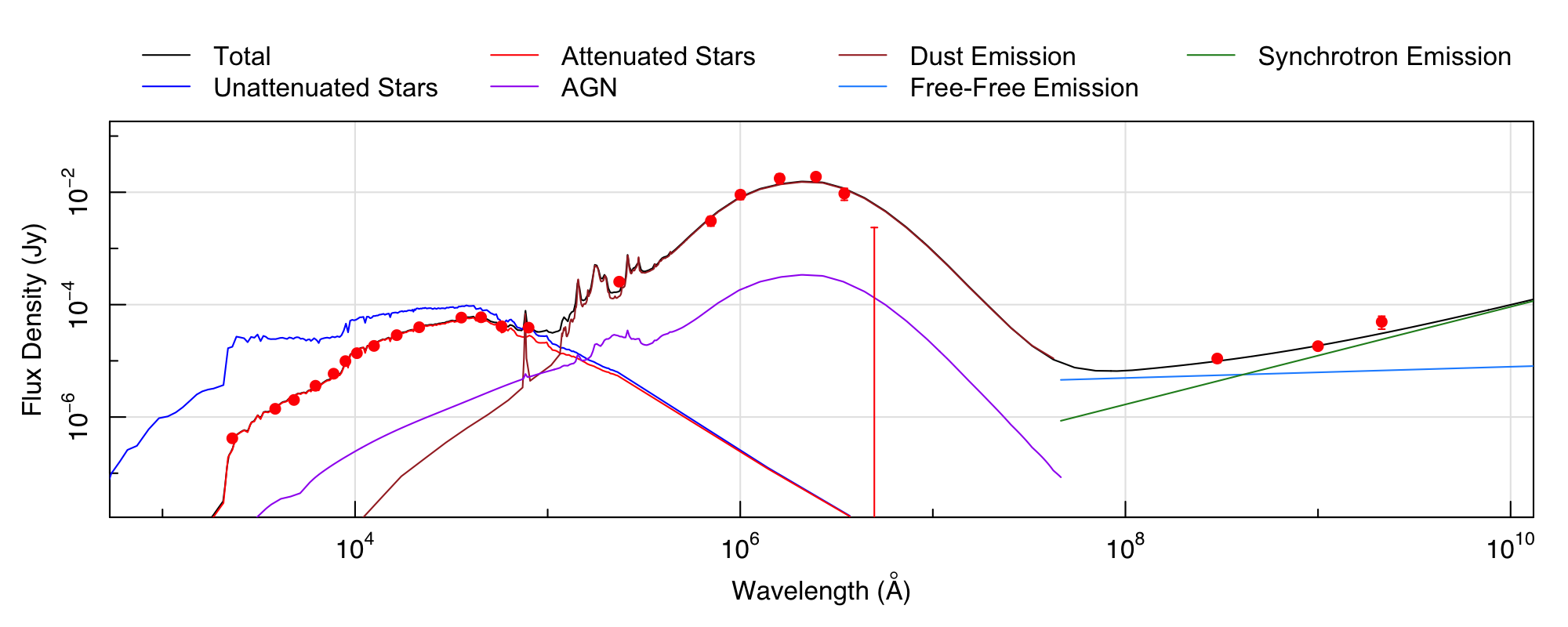}
    \caption[FUV-radio SED for DEVILS galaxy 101500680582617952]{The resulting FUV--radio SED fit for DEVILS galaxy 101500680582617952. 
    We show the input flux densities as the red points and error bars, and the total SED in black. 
    The contributing components are also shown including the unattenuated stellar emission (blue), attenuated stellar emission (red), AGN component (purple), dust emission (brown), free-free emission (light blue), and synchrotron emission (green).}
    \label{fig:RadioSED}
\end{figure*}

\begin{figure*}
    \centering
    \includegraphics[width=\linewidth]{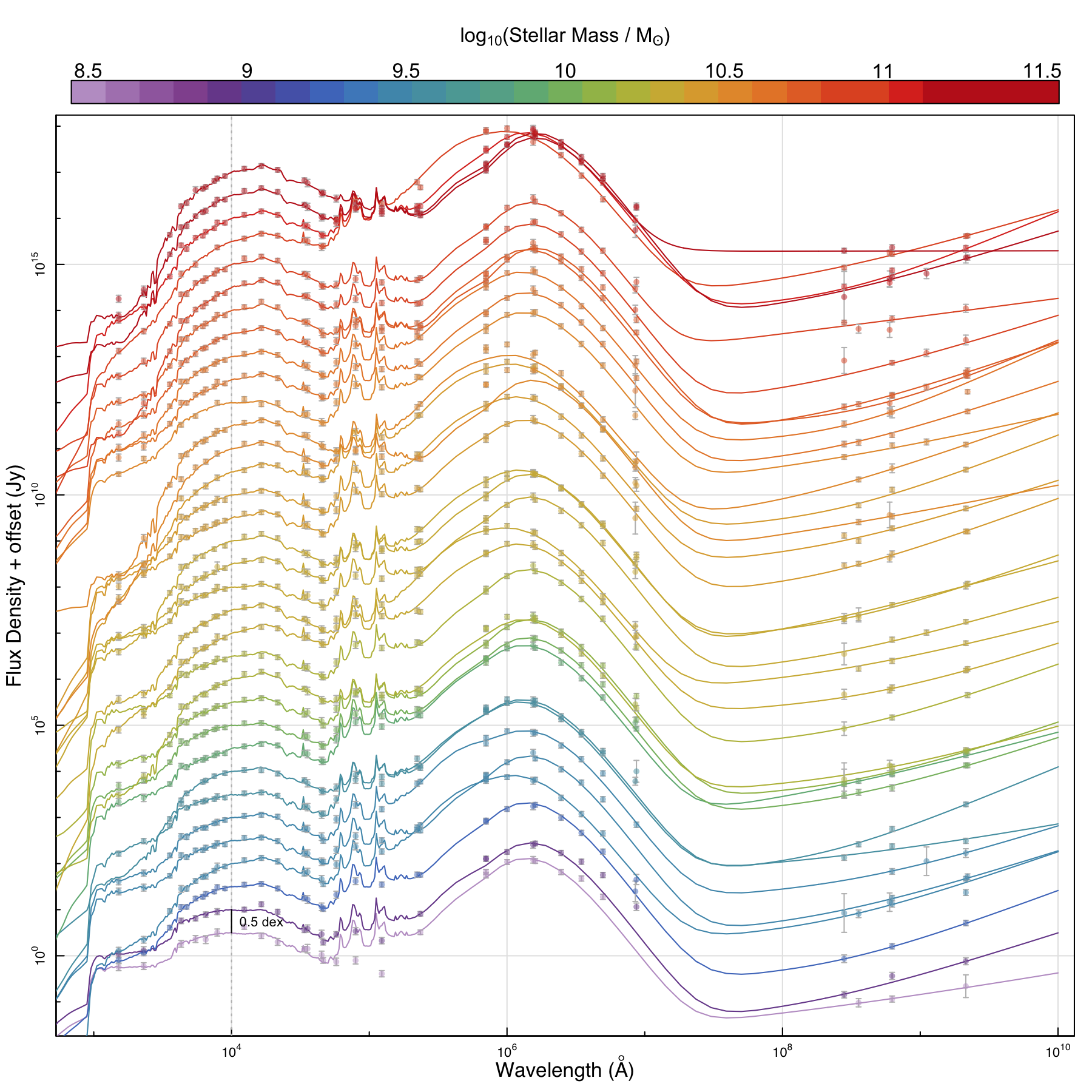}
    \caption[The input photometry and resulting \textsc{ProSpect} fits for the sample of KINGFISHER galaxies sorted by stellar mass.]{
    The input photometry (circles) and resulting \textsc{ProSpect} fits for the sample of KINGFISHER galaxies sorted from least (purple) to most (red) massive as per the colour bar. 
    To highlight the differences in SED between each galaxy we normalise each SED by the flux density at 10,000\,\AA\ and apply a fixed 0.5\,dex offset.}
    \label{fig:KFStackedSEDs}
\end{figure*}

\subsection{Radio Extension}
\textsc{ProSpect} models the emission of a galaxy in radio frequencies using a mixture of free-free (thermal) and synchrotron (non-thermal) radio emission using a prescription based on the work of \cite{MarvilIntegratedRadioContinuum2015}.

Essentially, the radio continuum is computed by combining the relations from \citet{KennicuttGlobalSchmidtLaw1998}\footnote{The derivation of this relation makes a number of assumptions including a \citet{SalpeterLuminosityFunctionStellar1955} IMF, solar abundances and that dust re-radiates all of the bolometric luminosity. 
Changes to the metallicity and IMF will therefore change the predicted free-free radio flux density however this is expected to only be a minor difference.} \new{and \citet{MurphyCalibratingExtinctionfreeStar2011} to obtain a relation for the optically thin free-free radio flux density, $S_\text{ff}(\nu)$:
\begin{align}
    S_\text{ff}(\nu) &= 1.4 \times 10^{10} \left( \frac{FIR}{\text{W m}^{-2}} \right) \times \left( \frac{T_e}{10^4 \text{\,K}} \right)^{0.45} \times \left( \frac{\nu}{\text{GHz}}\right)^{\alpha_\text{ff}},
\end{align}
which, assuming $\alpha_\text{ff} = -0.1$, gives a normalisation at 1.4\,GHz of
\begin{align}
    S_\text{ff}(1.4\,\text{GHz}) &= 13536775000  \left( \frac{FIR}{\text{W m}^{-2}} \right) \times \left( \frac{T_e}{10^4 \text{\,K}} \right)^{0.45}
\end{align}
This can therefore be used the define the free-free contribution to the radio continuum as a function of frequency:
\begin{align}
    S_\text{ff}(\nu) &= S_\text{ff}(1.4\,\text{GHz})  \times \left(\frac{\nu}{1.4 \text{\,GHz}} \right)^{\alpha_\text{ff}}.
\end{align}}
In each case, we assume an electron temperature ($T_e$) of $10^4$\,K as per \citet{MarvilIntegratedRadioContinuum2015} \new{and a free-free spectral index ($\alpha_\text{ff}$) of -0.1}.
The \textit{FIR} parameter is used as an estimate of the FIR flux density between 42.5 and 122.5\,$\mu$m and is calculated in \citet{MarvilIntegratedRadioContinuum2015} by combining the 60 and 100\,$\mu$m flux densities into a single value.
Within \textsc{ProSpect} we instead use the \citet{DaleTwoParameterModelInfrared2014} dust emission model fits to calculate the total FIR emission associated with star formation.

The contribution to the radio from synchrotron emission is then calculated using the \new{free-free fraction ($f_\text{ff}$) and synchrotron spectral index (\asy) parameters}:
\begin{align}
    S_\text{sy} (\nu) &= \left( S_\text{ff}(1.4\,\text{GHz}) \times \frac{1 - f_\text{ff}}{f_\text{ff}} \right) \times \left( \frac{\nu}{1.4\,\text{GHz}} \right)^{\alpha_\text{sy}}
\end{align}

The total radio continuum flux density is then:
\begin{align}
    S_\text{tot} (\nu) &= S_\text{ff}(\nu) +  S_\text{sy} (\nu) 
\end{align}

For this work, we only consider radio emission associated with star formation (using \texttt{addradio\_SF}) as although some of our galaxies are expected to have radio emission produced by an AGN, this additional emission can be modelled by the synchrotron component. 
In these cases, the free-free fraction will be lower than expected from purely star formation and any changes to the slope of the radio continuum can be captured by the free synchrotron spectral index parameter. 
Radio emission associated with an AGN component can also be added in \textsc{ProSpect} using \texttt{addradio\_AGN} where the \textit{FIR} parameter is calculated using only the AGN component's contribution to the FIR emission.
However, modelling the radio continuum using four distinct power-law components would result in degenerate solutions unless significantly more data over a larger frequency range are used. 
 
In addition to the free parameters described above, \new{we allow the free-free fraction (\texttt{ff\_frac\_SF}) and synchrotron spectral index (\texttt{sy\_power\_SF}) parameters to vary freely but fix the free-free spectral index,\texttt{ff\_power\_SF} = -0.1.
The spectral index of the synchrotron radio emission is typically between -0.8 and -0.6 for synchrotron emission generated from star-formation \citet{CondonRadioemissionnormal1992}, however, when fitting, we allow it to vary freely between -2.2 and 0.
This allows for where the shape of the radio SED is flatter due to contributions from radio-loud AGN (this is also the range used by \citealt{TabatabaeiRadioSpectralEnergy2017}). 
The fraction of free-free H\textsc{II} nebular plasma free-free radio emission at 1.4\,GHz is modelled within \textsc{ProSpect} using the \texttt{ff\_frac\_SF} parameter.
\cite{CondonNewStarburstModel1990} suggest that the \new{free-free fraction} is typically seen to span the range 0.05 to 0.2 for star-forming galaxies.
Again, because some galaxies might have radio contributions from an AGN, we fit using a range of $\log_{10} (\texttt{ff\_frac\_SF}) = [-5,-0.5]$, where a lower value corresponds to less free-free emission, and more contribution from synchrotron emission either from star-formation or an AGN. }
Both radio parameters are fit simultaneously with the usual 12 parameters relating to the star formation and metallicity histories, and the dust and AGN components.

Table~\ref{tab:ParametersRadio} lists all the parameters used in this work, a brief description, whether they are fit in linear or logarithmic space or are fixed, and the allowed values.

\new{
We implement \textsc{ProSpect} in a Bayesian manner using a combination of genetic optimisation and Markov Chain Monte Carlo (MCMC) phases using the \textsc{Highlander} R package\footnote{\url{https://github.com/asgr/Highlander}}.
\textsc{Highlander} alternates between genetic optimization using the \textsc{cmaeshpc}\footnote{\url{https://github.com/asgr/cmaeshpc}} and an MCMC chain using the \textsc{LaplacesDemon}\footnote{\url{ https://cran.r- project.org/web/packages/LaplacesDemon/index.html}} package.
By alternating between the two different phases, \textsc{Highlander} is able to more efficiently sample the posterior parameter space, especially in scenarios that are highly multimodal, such as SED fitting, whilst still retaining the ability to extract uncertainties for each of the galaxy properties. 
Within \textsc{LaplacesDemon}, we utilize the CHARM\footnote{Component-wise hit and run metropolis.} algorithm using a Student-t likelihood. 
We fit with 2000 steps for both the genetic optimisation and MCMC phases, repeating each phase twice for a total of 8000 iterations. 
The values quoted within this work correspond to the maximum likelihood step and uncertainties are taken to be the 1$\sigma$ range of the final MCMC phase. 
We include a 10 per cent error floor \new{added in quadrature} on all measurements to account for offsets between facilities and instruments. 
}

\begin{figure*}
    \centering
    \includegraphics[width = \linewidth]{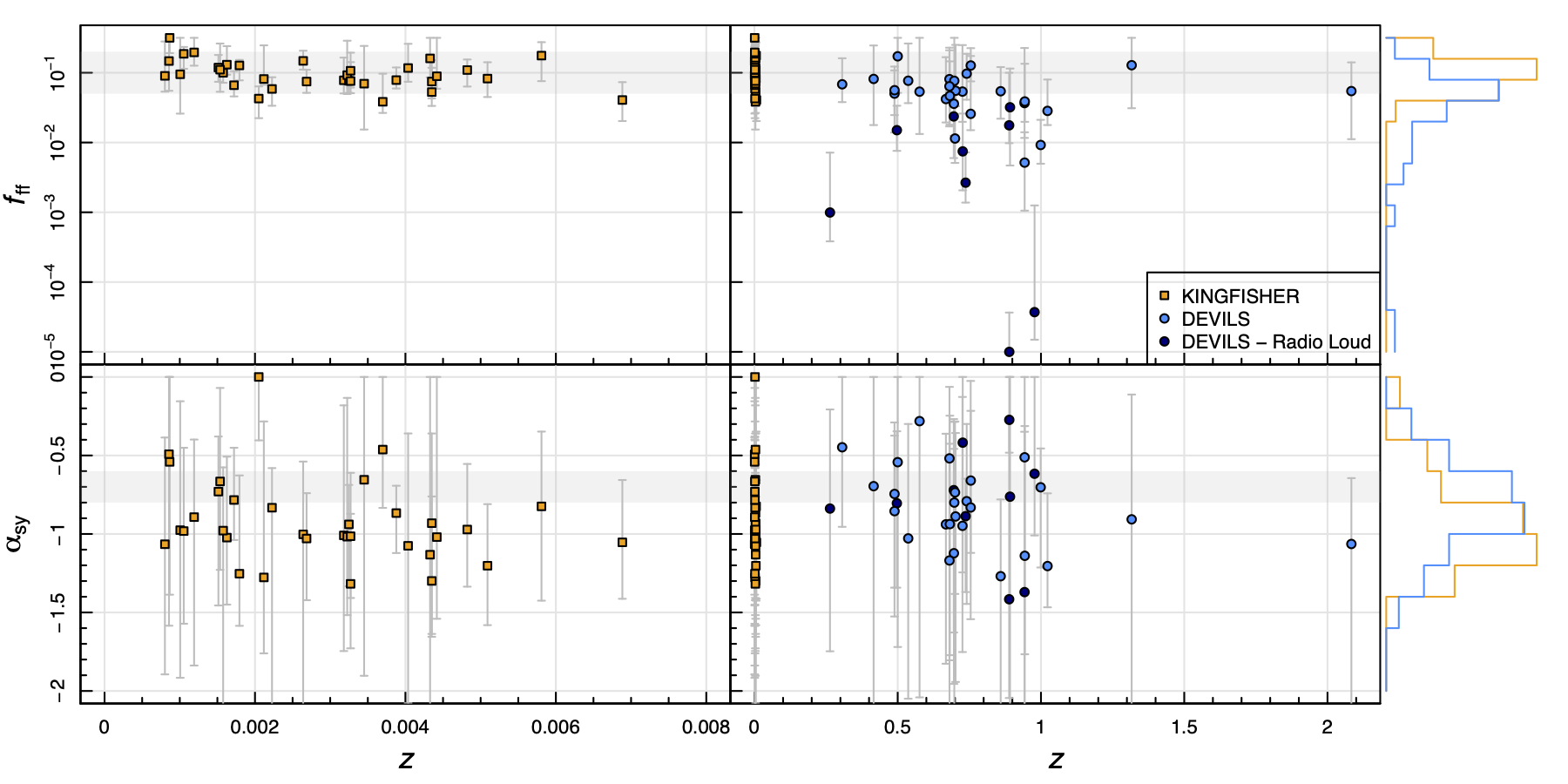}
    \caption[The distribution of recovered free-free fraction (\fff) and synchrotron spectral index (\asy) parameters as a function of redshift.]{The distribution of recovered free-free fractions (\fff) and synchrotron spectral index (\asy) parameters as a function of redshift. 
    The grey shaded region shows the expected ranges for the two parameters as per \citet{CondonRadioemissionnormal1992}.
    We show the sample of KINGFISHER galaxies as the orange squares and the DEVILS sample as the blue circles.
    For the DEVILS sample, we show galaxies identified as radio-loud AGN by \citet{WhittamMIGHTEEnatureradioloud2022} as the navy blue circles.
    The right panel shows the full redshift range, while the left panel shows a zoom of the range covered by the KINGFISHER sample. 
    \new{The histograms on the outer right edge show the projected density of the recovered free-free fractions and synchrotron spectral indices for the two samples.}
    }
    \label{fig:RadioRedshiftTrends}
\end{figure*}

\subsection{\textsc{ProSpect} Outputs}
Figure~\ref{fig:RadioSED} shows the input FUV-radio photometry for an example DEVILS galaxy, as well as the resulting \textsc{ProSpect} fit and contributions from the stellar, dust, AGN, and radio free-free and synchrotron emission. 
In this case, the free-free emission is the dominant source of radio emission at 10\,GHz and results in a bending in the radio-SED that would not be accounted for if fitting a single power law as in \citet{DeyLowfrequencyRadioContinuum2022}. 

\new{The resulting SED for each galaxy is visually inspected to verify that the fit is appropriate and in general, the resulting SED fits are good especially for the KINGFISHER sample due to the greater photometric coverage.}
In addition, when converting the likelihoods to $\chi^2$ values, all galaxies across DEVILS and KINGFISHER have a reduced $\chi^2 < 4$.
For the DEVILS sample, the fits in the FIR regime are less constrained than the KINGFISHER galaxies due to lower SNR photometry. 
One such galaxy has radio emission at higher flux densities than the FUV-MIR and FIR photometry mostly consisting of upper limits. 
This causes a discontinuity in the best-fitting SED at the point where the dominant emission mechanism shifts from the non-thermal radio continuum to emission from dust. 
This particular galaxy is classed as an RLAGN in the catalogue from \citet{WhittamMIGHTEEnatureradioloud2022} and radio jets can be seen in the 1.4\,GHz image.
This explains the very bright radio emission relative to the host galaxy.

Figure~\ref{fig:KFStackedSEDs} shows the resulting SEDs for the sample of KINGFISHER galaxies sorted by stellar mass (each of the individual SEDs, component contributions, and resulting star formation and metallicity histories are made available as supplementary material). 
This Figure highlights the fact that the diversity in radio SEDs is observationally driven and is not just an unconstrained extension to the \textsc{ProSpect} model.

\section{Radio SED Parameters} \label{sec:RadioSEDParameters}

\begin{figure}
    \centering
    \includegraphics[width =\linewidth]{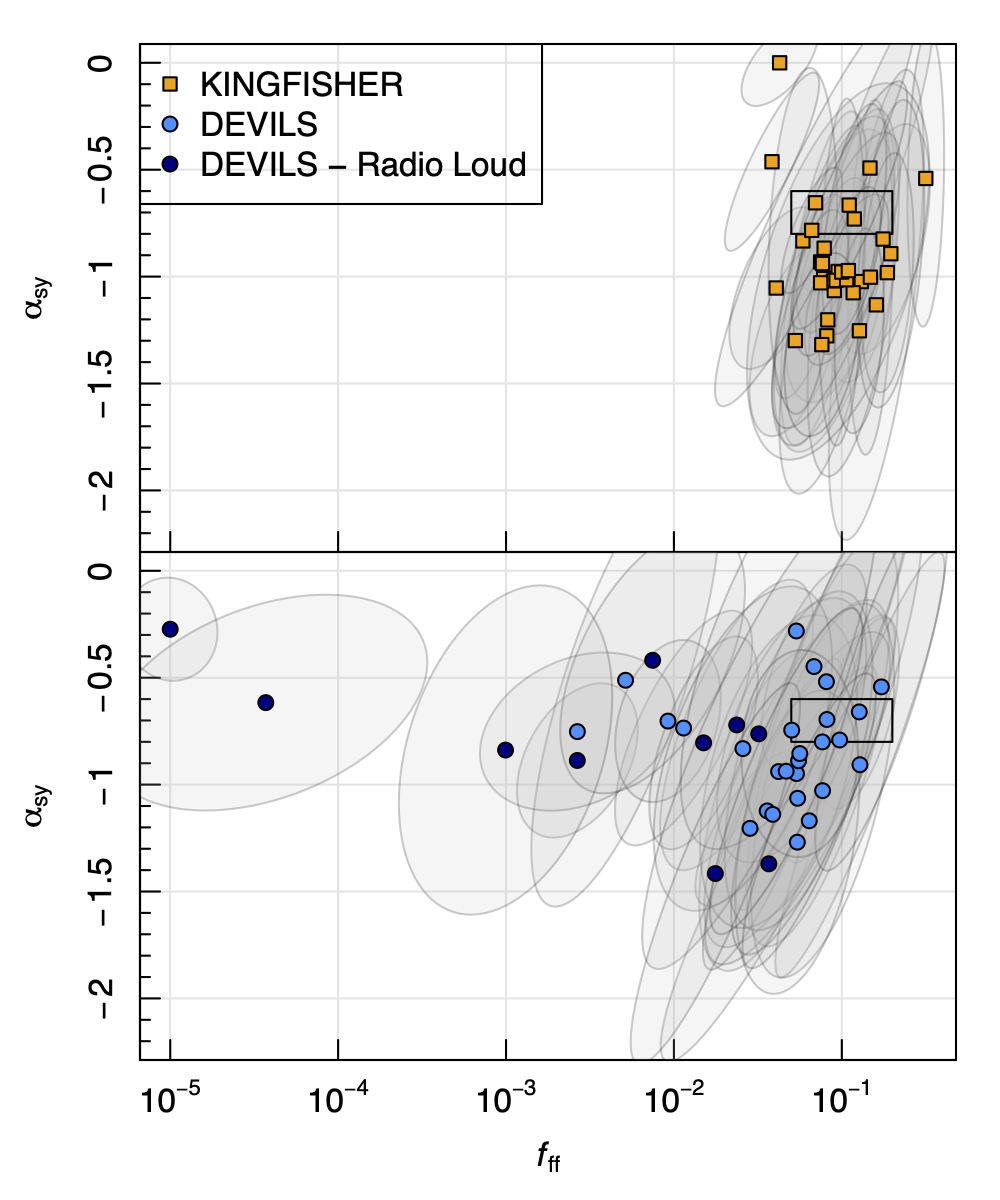}
    \caption[The synchrotron spectral index (\asy) against the free-free fraction at 1.4\,GHz (\fff)]{
    The synchrotron spectral index (\asy) against the free-free fraction at 1.4\,GHz (\fff) for the KINGFISHER (upper) and DEVILS samples (lower). 
    The ellipses show the correlation between the two parameters in the posterior chains.
    The black rectangle shows the expected range of the two parameters from \citet{CondonRadioemissionnormal1992}. }
    \label{fig:RadioParams}
\end{figure}

\begin{figure*}
    \centering
    \includegraphics[width = \linewidth]{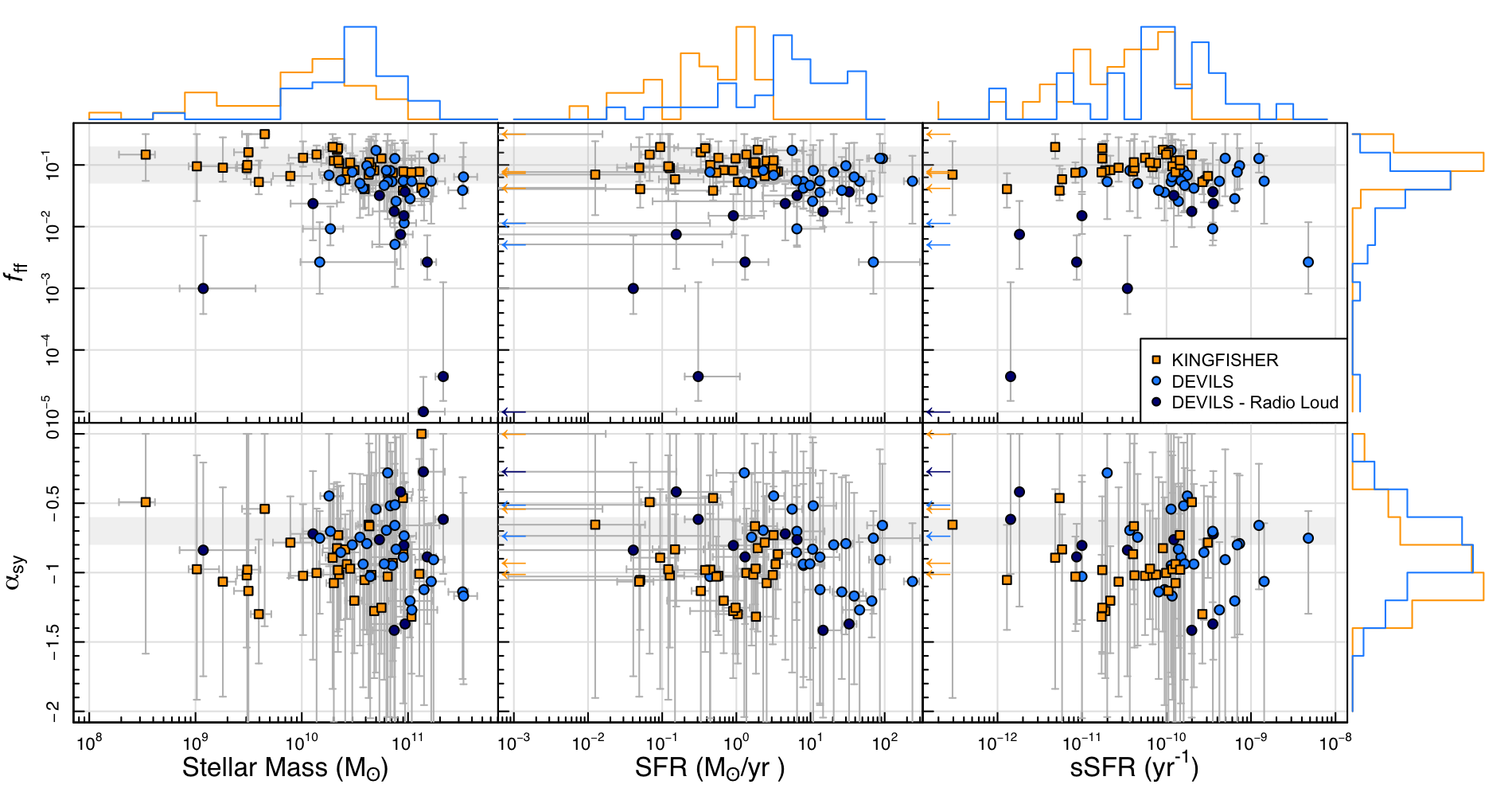}
    \caption[The distribution of recovered free-free fractions and synchrotron spectral index parameters as a function of stellar mass, SFR, and sSFR.]{The distribution of recovered free-free fractions (\fff) and synchrotron spectral index (\asy) parameters as a function of stellar mass (left), SFR (middle), and sSFR (right).
    The colouring of points and grey shaded region are as per Figure~\ref{fig:RadioRedshiftTrends}.
    \new{The outer histograms show the projected distribution of the KINGFISHER (orange) and DEVILS (blue) samples for each property. }
    }
    \label{fig:RadioStellarTrends}
\end{figure*}

\begin{figure*}
    \centering
    \includegraphics[width = \linewidth]{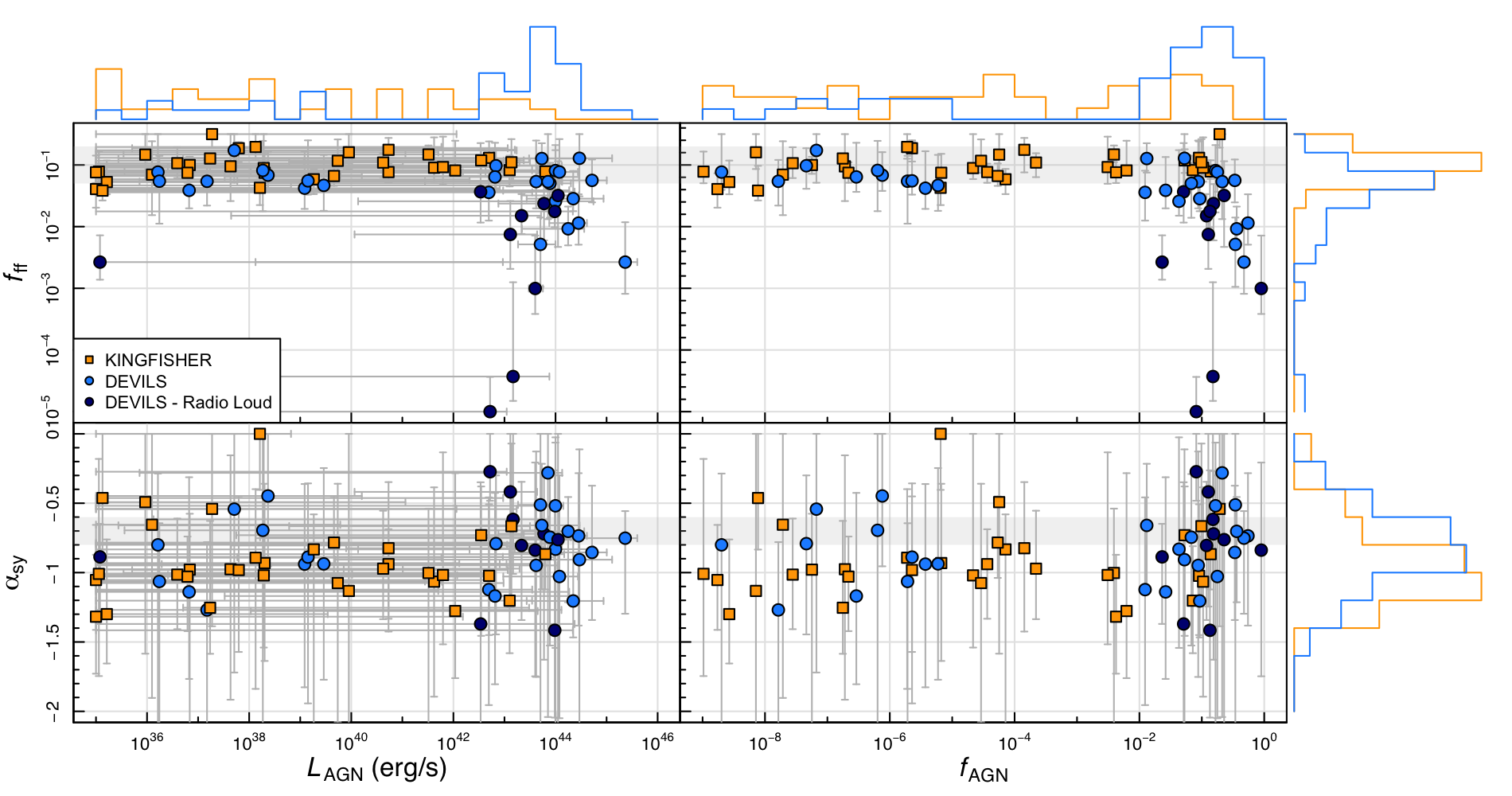}
    \caption[The distribution of recovered free-free fractions and synchrotron spectral index parameters as a function of AGN luminosity (left) and $f_\text{AGN}$ (right).]{The distribution of recovered free-free fractions (\fff) and synchrotron spectral index (\asy) parameters as a function of AGN luminosity (left) and $f_\text{AGN}$ (right).
    The colouring of points and grey shaded region are as per Figure~\ref{fig:RadioRedshiftTrends}.
     \new{The outer histograms show the projected distribution of the KINGFISHER (orange) and DEVILS (blue) samples for each property. }}
    \label{fig:RadioAGNTrends}
\end{figure*}

The distributions of the \textsc{ProSpect} parameters arising from the FUV-FIR portion of SEDs have been extensively explored in previous work \citep{BellstedtGalaxyMassAssembly2020b,BellstedtGalaxyMassAssembly2021,ThorneDeepExtragalacticVIsible2021,ThorneDeepExtragalacticVIsible2022,ThorneDEVILScosmicevolution2022}. 
However, to validate the implementation of the radio extension, this work explores the resulting radio-parameters and their correlations with other galaxy properties.

Figure~\ref{fig:RadioRedshiftTrends} shows the distribution of the derived free-free fractions ($f_\text{ff}$) and synchrotron spectral indices ($\alpha_\text{sy}$) as a function of redshift for our sample.
Due to the very different redshift ranges sampled by the KINGFISHER and DEVILS samples, we show the distribution of just the KINGFISHER galaxies in the left panel, while the right panel shows the redshift distribution of both samples. 
For the DEVILS sample, we recover a median $f_\text{ff} = 0.05 \pm 0.007$ and $\alpha_\text{sy} =-0.84 \pm 0.05$. 
\new{When removing the 14 galaxies identified as AGN} by \citet{WhittamMIGHTEEnatureradioloud2022} or \textsc{ProSpect} from the DEVILS sample, the median free-free fraction of the remaining 19 galaxies increases slightly to $f_\text{ff} = 0.055 \pm 0.007$ while the median synchrotron spectral index steepens slightly ($\alpha_\text{sy}=-0.9 \pm 0.07$) \new{however, within errors these are still consistent with the full sample}.
\new{For the sample of KINGFISHER galaxies, we recover a slightly higher median $f_\text{ff} = 0.09 \pm 0.009$ than the DEVILS sample and a synchrotron \new{spectral index} $\alpha_\text{sy} =-0.98 \pm 0.03$. }
Our median free-free fraction for the KINGFISHER sample is consistent with those from surveys of nearby galaxies (8-10 per cent; \citealt{Kleinsurveydistributionscm1981,Kennicuttoriginnonthermalradio1983,Niklasradiocontinuumsurvey1997}).
However, the median free-free fraction of the DEVILS sample is lower than previous estimations due to our sample selection not excluding AGN and could indicate that the RLAGN selection from \citet{WhittamMIGHTEEnatureradioloud2022}  could be missing lower luminosity RLAGN. 
We recover a steeper synchrotron \new{spectral index} than the canonically assumed value of $\alpha \sim -0.8$ \citep{CondonRadioemissionnormal1992} and also steeper than the value derived by \citet{Kleinsurveydistributionscm1981,Niklasradiocontinuumsurvey1997} as they included frequencies below 1\,GHz where free-free absorption can cause a flattening of the SED \new{\citep{LackiInterpretinglowfrequencyradio2013,CalistroRiveraLOFARwindowstarforming2017}}.
However, our synchrotron \new{spectral index} is closer to that derived by \citet{Galvinspectralenergydistribution2018}, $\alpha_{sy} = -1.06$, who model the 70\,MHz–48\,GHz SEDs of 19 starburst galaxies and account for free-free emission and self-absorption as well as synchrotron emission. 
Our \fff\ and \asy\ values are also in agreement with those from \citet{TabatabaeiRadioSpectralEnergy2017} who also use the KINGFISHER sample and recover a mean $\alpha_\text{nt} = 0.97 \pm 0.16$ and $f_\text{th} = (10 \pm 9)$ per cent at 1.4 GHz. 

We find no clear evolution in the free-free fraction or synchrotron \new{spectral index} as a function of redshift.
However, we find that the median free-free fraction of the DEVILS sample is at the lower limit of the expected range from \citet{CondonRadioemissionnormal1992} and is lower than the median of the KINGFISHER sample. 
\new{Of the DEVILS galaxies, we recover maximum likelihood free-free fractions below 0.05 for 16 of the 33 galaxies, with four of these galaxies with free-free fractions more than $3\sigma$ below $f_\text{ff} = 0.05$.
For the KINGFISHER sample, only three of the 34 galaxies have a best-fitting free-free fraction below 0.05, however, in all three cases, the  $1\sigma$ uncertainty is consistent with $f_\text{ff}=0.05$}
This trend is most likely driven by the differing selection of the two samples. 
While the KINGFISH sample includes galaxies with an AGN nuclear type as derived from optical emission line diagnostic plots, these are all low-luminosity AGN except for NGC 1316 (Fornax A) which was not included in the radio observations from \citet{TabatabaeiRadioSpectralEnergy2017} and is therefore not included in our final sample. 
On the other hand, our DEVILS sample selection did not remove AGN. 
Radio-loud AGN would be expected to have lower free-free fractions, therefore lowering the median \fff\ value for the DEVILS sample. 
This is evident in the DEVILS source with \fff\ $=10^{-5}$ \new{which lies at the lower limit of the allowed range} and is classified as a radio-loud AGN based on the IRRC.  

Figure~\ref{fig:RadioParams} shows the two radio parameters plotted against each other with the ellipses showing the correlation from the final posterior chain. 
As expected, for each galaxy the \fff\ and \asy\ parameters are strongly correlated (evident as the diagonal trend in the ellipses) due to degeneracies when fitting where a flattening of the synchrotron spectrum is associated with an increase in free-free fraction at 1.4\,GHz.The correlation between \fff\ and \asy\ on a galaxy-by-galaxy basis is seen for both the KINGFISHER and DEVILS samples, however, the overall population values are not strongly correlated.

When fitting just the radio portion of a galaxy's SED, the free-free fraction is difficult to derive as it relies on fitting over a wide frequency range to recover the change in slope at high frequencies from the increasing contribution from free-free emission. 
However, when simultaneously fitting the FUV-FIR SED, the value of normalisation of the free-free emission, and therefore the value of \fff, can be more easily derived as it is coupled to the SFR and dust properties of the galaxy. 
The \asy\ is only constrained by the radio measurements which are often impacted by differences between resolution, source detection, and flux density extraction techniques. 
These differences can lead to systematic flux density offsets between frequencies and therefore to poorly constrained spectral indices (see Appendix~\ref{app:radioSEDs} for the impact of the large \asy\ uncertainty on the resulting SED). 

Figure~\ref{fig:RadioStellarTrends} shows the distribution of the radio parameters as a function of stellar mass, SFR, and sSFR.
In each case, we find no clear trend with either \fff\ or \asy. 
We find that the derived free-free fractions are typically in better agreement with the suggested values from \citet{CondonRadioemissionnormal1992} than the synchrotron \new{spectral indices} which show a large scatter within the allowed range. 

Figure~\ref{fig:RadioAGNTrends} shows the distribution of derived free-free fractions and synchrotron \new{spectral indices} as a function of the bolometric AGN luminosity ($L_\text{AGN}$) and $f_\text{AGN}$. 
When considering the distribution of recovered free-free fractions we find that most objects with practically no MIR contribution from an AGN ($f_\text{AGN} < 0.01$) have a free-free fraction consistent with expected values for `normal' galaxies (0.05-0.2; \citealt{CondonRadioemissionnormal1992}). 
We also find that the radio-loud AGN all have a \textsc{ProSpect}-derived $f_\text{AGN} > 0.02$, with a median $f_\text{AGN} = 0.13$.
These objects also have \fff $< 0.04$ corresponding to more radio emission than would be expected from star-formation alone (as per \citealt{CondonRadioemissionnormal1992}).

As our \textsc{ProSpect} fits recover radio parameters (\fff\ and \asy) consistent with previous studies, and the free-free fractions of RLAGN behave as expected, we can be confident that our implementation does not obviously bias the recovery of radio parameters when simultaneously fitting FUV-radio SEDs. 

\subsection{Comparisons to Tabatabaei et al. (2017)}
\begin{figure}
    \centering
    \includegraphics[width = \linewidth]{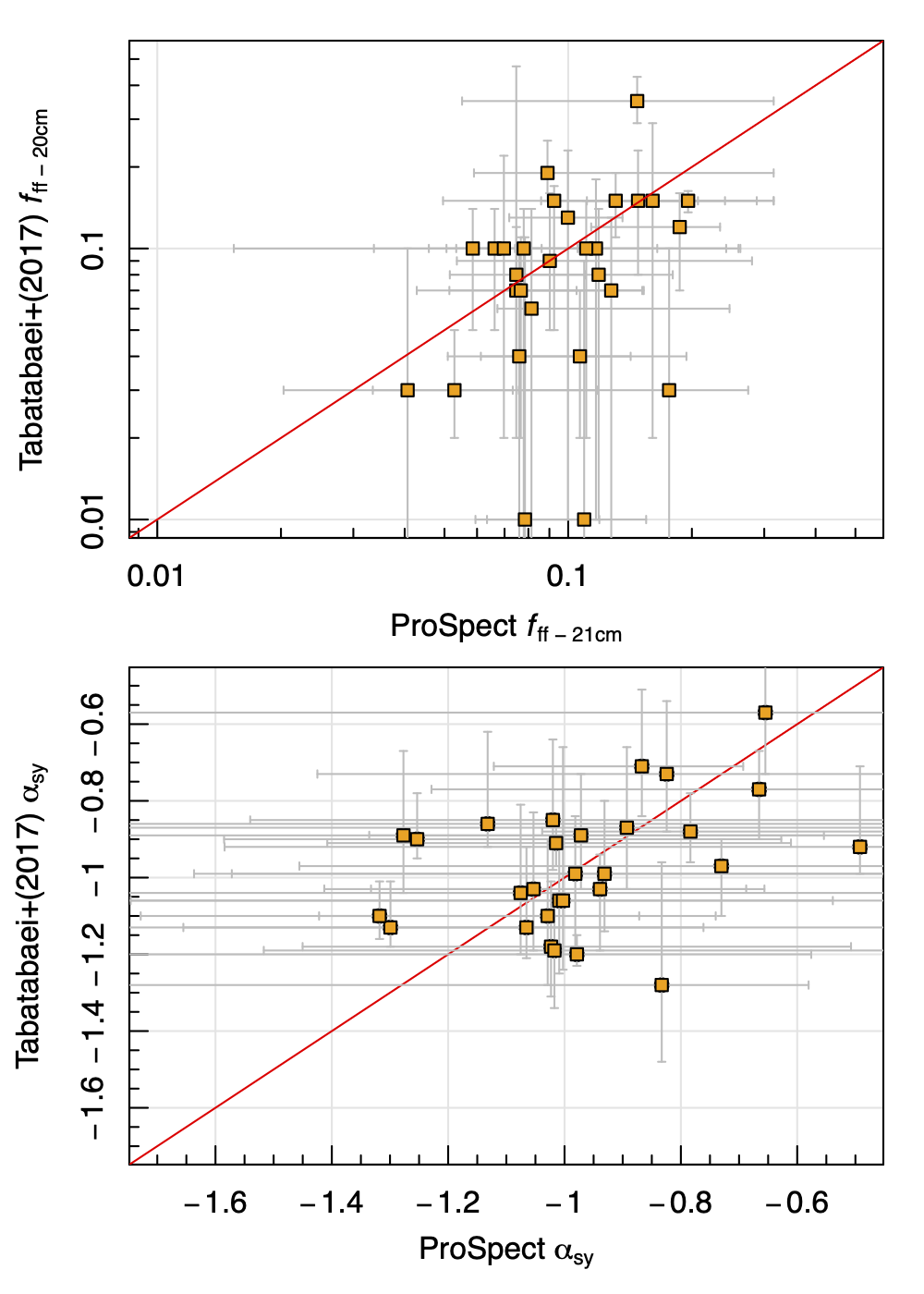}
    \caption[Comparisons of the derived free-free fractions and synchrotron \new{spectral indices} with those from \citet{TabatabaeiRadioSpectralEnergy2017}]{Comparisons of the derived free-free fractions and synchrotron \new{spectral indices} as derived by \citet{TabatabaeiRadioSpectralEnergy2017} with those derived in this work for the KINGFISHER sample.
    The red line shows the one-to-one relation.
}
    \label{fig:TabatabaeiComparisons}
\end{figure}

\begin{figure}
    \centering
    \includegraphics[width = \linewidth]{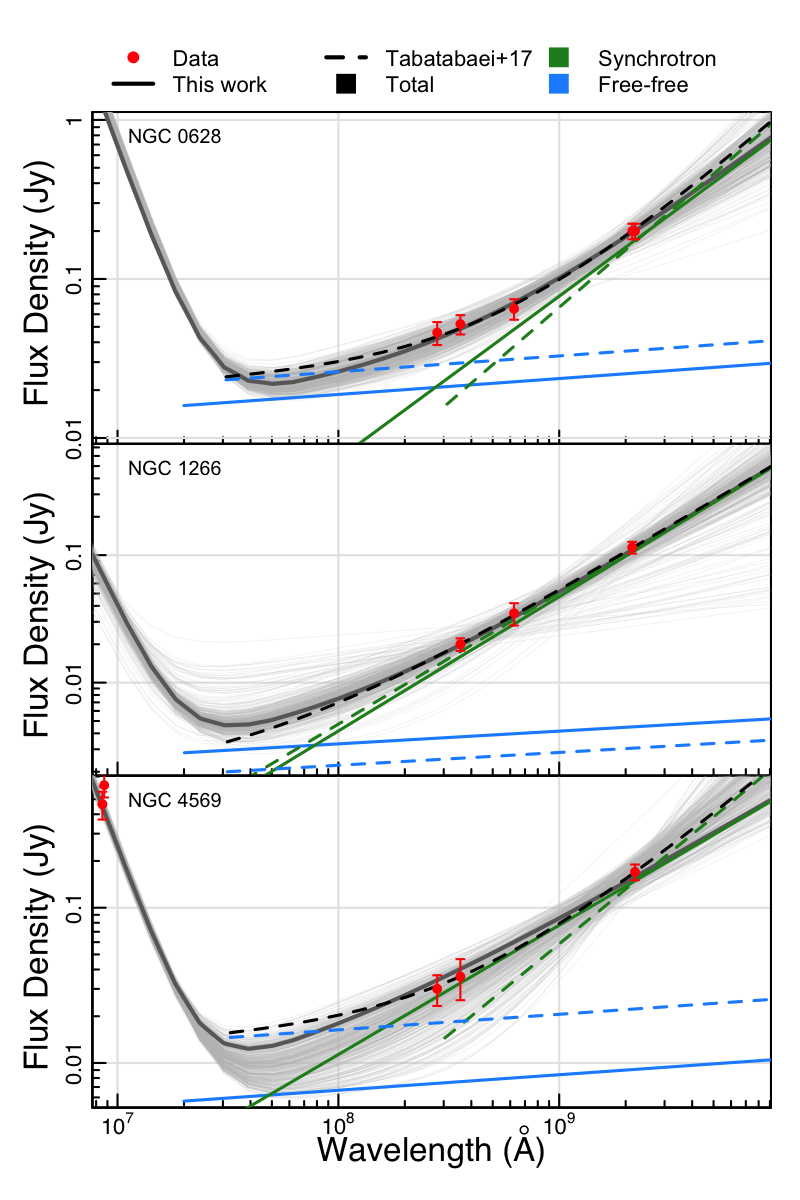}
    \caption[Comparisons of the radio SEDs obtained using \textsc{ProSpect} with those from \citet{TabatabaeiRadioSpectralEnergy2017}.]{Comparisons of the radio SEDs for three example KINGFISHER galaxies (NGC 0628, NGC 1266, NGC 4569)  obtained using \textsc{ProSpect} with the fits from \citet{TabatabaeiRadioSpectralEnergy2017}. 
    In each panel we show the input data (red points), both the free-free (blue) and synchrotron (green) contributions to the total radio continuum.
    In all cases, both the \textsc{ProSpect} and \citet{TabatabaeiRadioSpectralEnergy2017} fits provide a good fit to the data. 
    }
    \label{fig:tabatabaeiSEDcomparisons}
\end{figure}

\begin{figure}
    \centering
    \includegraphics[width = \linewidth]{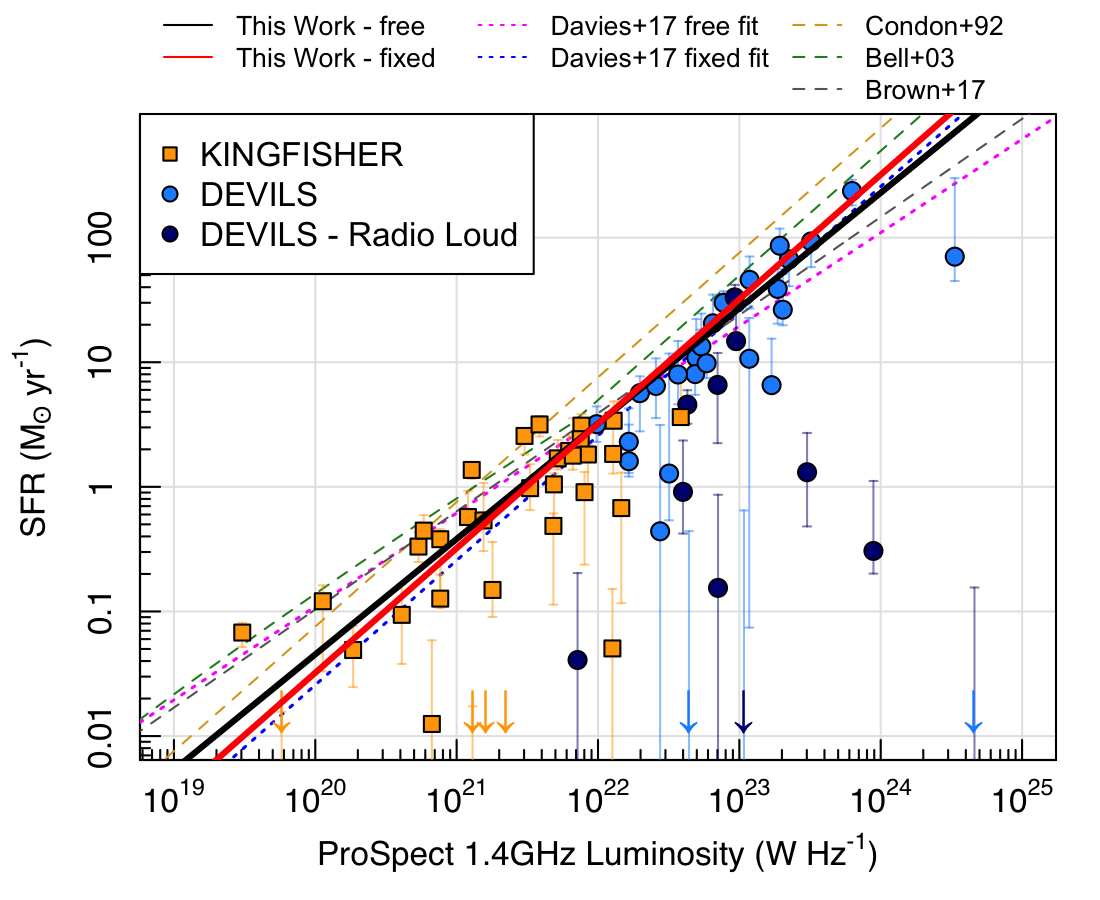}
    \caption[Correlation between 1.4\,GHz luminosity and \textsc{ProSpect}-derived SFR.]{Correlation between 1.4\,GHz luminosity and \textsc{ProSpect}-derived SFR for the KINGFISHER (orange) and DEVILS (blue) samples. 
    The dark blue indicates galaxies that were classed as RLAGN by \citet{WhittamMIGHTEEnatureradioloud2022}. 
    We show our linear fits with the free and fixed slopes as the black and red solid lines respectively.
    The free and fixed fits from \citet{DaviesGalaxyMassAssembly2017} are also shown as the dotted magenta and blue lines. 
    We also show the relations from \citet{CondonRadioemissionnormal1992},\citet{BellEstimatingStarFormation2003} and \citet{BrownCalibrationultravioletmidinfrared2017} as the dashed yellow, green, and grey lines respectively. }
    \label{fig:SFR_1.4GHzLum}
\end{figure}

The SEDs of the KINGFISHER sample have been fit previously, however the FUV-FIR and radio regimes have only been considered separately. 
The radio SEDs of the KINGFISHER galaxies were previously fit by \citet{TabatabaeiRadioSpectralEnergy2017} while the FUV-FIR SEDs were fit by \citet{HuntComprehensivecomparisonmodels2019}. 
In this Section, we compare the estimates for the radio free-free fractions (\fff) and synchrotron \new{spectral indices} (\asy) as derived by \textsc{ProSpect} to the results from \citet{TabatabaeiRadioSpectralEnergy2017}.
We compare the results for other galaxy properties to those using \textsc{magphys}, \textsc{cigale}, and \textsc{grasil} from \citet{HuntComprehensivecomparisonmodels2019} in Appendix~\ref{app:KINGFISHERSEDComparisons}.
As discussed in \citet{ThorneDeepExtragalacticVIsible2021} our flexible parametric SFH and evolving metallicity history implementation recovers systematically higher stellar masses than other codes that use variants of exponentially declining SFHs. 
We find that our dust luminosities are consistent with all three fitting techniques used by \citet{HuntComprehensivecomparisonmodels2019} however the recovered dust masses are highly dependent on the assumed mass to light conversions.

To fit the radio SEDs of the KINGFISHER galaxies, \citet{TabatabaeiRadioSpectralEnergy2017} use two power-law components to model the contribution from the free-free and synchrotron emission. 
Similarly to our \textsc{ProSpect} fits, they also fix the slope of the \new{free-free spectral index to -0.1 and fit for the synchrotron spectral index} and the normalisations of the two components. 
The normalisations for each component were allowed to take negative values to test the robustness of the fit and to assess whether a free-free component was required. 
The two normalisation terms can be combined into a free-free fraction estimate at any frequency however we compare to the fractions derived at 20\,cm (1.49\,GHz).
For five galaxies (IC0342, NGC 1482, NGC 3077, NGC 4236, and NGC 4579), \citet{TabatabaeiRadioSpectralEnergy2017} found that two-component fits resulted in negative free-free fractions and so a single component was preferred. 
We do not include these galaxies in Figure~\ref{fig:TabatabaeiComparisons}. 

Figure~\ref{fig:TabatabaeiComparisons} shows the comparison of the two radio parameters derived with \textsc{ProSpect} with the results from \citet{TabatabaeiRadioSpectralEnergy2017}. 
We find reasonable agreement between the free-free fractions, with a few objects offset by $>0.5$\,dex (NGC 2146, NGC 2798, NGC 3265, NGC 5713) however, within errors, these are consistent.
Despite using the same allowed ranges, our \asy\ values have a larger scatter when compared to \citet{TabatabaeiRadioSpectralEnergy2017} than the free-free fractions.
However, the derived \new{spectral indices} are relatively consistent with a median absolute offset of 0.17 with 90 per cent of the sample offset by less than 0.3. 

We also show comparisons of the resulting fits in Figure~\ref{fig:tabatabaeiSEDcomparisons} for three of the KINGFISHER galaxies (NGC 0628, NGC 1266, NGC 4569). 
The galaxies were selected to show a case where the \new{best fitting radio parameters are in good agreement (NGC 1266), a case where they are in reasonable agreement (NGC 0628), and a case with the largest difference in synchrotron \new{spectral index} (NGC 4569).
However, in all cases, the derived radio parameters are within errors} and both the \textsc{ProSpect} and \citet{TabatabaeiRadioSpectralEnergy2017} fits provide a good fit to the data. 
Significant differences between the best-fitting total SEDs are not apparent within the frequency range examined here and determining the true shape of the radio SED would require additional measurements at higher frequencies.

These comparisons demonstrate that, even with the added complexity of simultaneously modeling the FUV-FIR, our derived radio parameters, and best-fitting radio SEDs are consistent with those derived using a simpler technique.

\section{1.4 GHz Luminosity -- SFR Relation}\label{sec:L14SFR}

Using the \textsc{ProSpect} fits for both the KINGFISHER and DEVILS samples we can also investigate the 1.4\,GHz luminosity--SFR relation.
The use of both DEVILS and KINGFISHER allows for a better sampling of the relation over a larger range of SFRs and luminosities, and to explore whether the relation evolves with redshift. 

Figure~\ref{fig:SFR_1.4GHzLum} shows the relation using the \textsc{ProSpect}-derived SFRs and the rest-frame 1.4\,GHz luminosity derived using the associated best-fitting SED for both samples. 
We highlight sources that have been classed as RLAGN by \citet{WhittamMIGHTEEnatureradioloud2022} as the navy blue points. 
We find that most of our sample follow a tight relation, where a higher 1.4\,GHz luminosity is associated with a higher SFR. 
However, we find that some galaxies lie below the relation. 
A number of these are identified as RLAGN by \citet{WhittamMIGHTEEnatureradioloud2022} or have very low SFRs indicative of passive systems (shown as the downward arrows). 
Interestingly, we find that the relationship is consistent between samples and shows no obvious discontinuity or change in slope between the KINGFISHER and DEVILS samples. 
This suggests no evolution in the relationship for $z<1.5$. 

We show previously published relations outlined in \citet{BellEstimatingStarFormation2003}, \citet{CondonRadioemissionnormal1992} \citet{BrownCalibrationultravioletmidinfrared2017} as the dashed green, yellow, and grey lines respectively. 
We convert these to a \citet{ChabrierGalacticStellarSubstellar2003} IMF using the conversions outlined in \citet{HaarsmaFaintRadioSources2000} for Miller-Scalo to Salpeter, and \citet{DriverTwophasegalaxyevolution2013} for Salpeter to Chabrier. 
We also show the relations from \citet{DaviesGalaxyMassAssembly2017}.
These were derived using a combination of GAMA and the Faint Images of the Radio Sky at Twenty-cm (FIRST, \citealt{BeckerFIRSTSurveyFaint1995}) surveys using both detections and stacking. 
The fits from \citet{DaviesGalaxyMassAssembly2017} shown in Figure~\ref{fig:SFR_1.4GHzLum} use \textsc{magphys}-derived SFRs and were performed using \textsc{hyperfit} using a fixed, $m=1$, slope (blue line), and a free slope and normalisation (magenta). 

Following the procedure from \citet{DaviesGalaxyMassAssembly2017} we also use \textsc{hyperfit} \citep{RobothamHyperFitFittingLinear2015} to fit the 1.4\,GHz luminosity -- SFR relation using a free slope and normalisation. 
We only include sources that are not identified as AGN by either \citet{WhittamMIGHTEEnatureradioloud2022} or \textsc{ProSpect} (i.e. removing sources with $f_\text{AGN} > 0.1$).
\new{We also remove the two KINGFISHER galaxies with very low SFRs (where the 99.9th percentile of the SFR posterior is $< 10^{-5}\,M_\odot \text{ yr}^{-1}$) that lie significantly off the relation. }

When fitting freely we derive a relationship between SFR and 1.4\,GHz luminosity given by:
\begin{multline}\label{eq:freeSFRL14}
    \log_{10} (\text{SFR} / M_\odot \,\text{yr}^{-1} ) = ( 0.926 \pm 0.08 ) \times \log_{10} (L_{1.4\,\text{GHz}} / \text{W\,Hz}^{-1})  \\ - (19.87 \pm 1.8) .
\end{multline}
We also fit assuming a unity slope which results in a best-fitting relation of 
\begin{equation}\label{eq:fixedSFRL14}
    \log_{10} (\text{SFR} / M_\odot \,\text{yr}^{-1} ) =  \log_{10} ( L_{1.4\,\text{GHz}} / \text{W\,Hz}^{-1} ) - (21.47 \pm 0.23) .
\end{equation}

Interestingly, within errors, the free and fixed slope fits are consistent. 
As free-free radio emission is known to scale linearly with SFR from the fundamental theory of emission processes \citep{CondonRadioemissionnormal1992}, this also suggests that synchrotron emission also scales linearly with SFR.
This is contrary to the results of \citet{DaviesGalaxyMassAssembly2017} who recover a best-fitting relation with $m=0.75 \pm 0.03$ suggesting that synchrotron emission does not scale linearly with star formation. 

Although the derivation of both the SFR and 1.4\,GHz rest-frame luminosity are linked, our findings are not a forced outcome of the model. 
The 1.4\,GHz luminosity is dependent on the relative contributions of the free-free and synchrotron radio emission and the slope of the synchrotron component which are both modeled as free parameters in our implementation.
Therefore, the relation given in Equation~\ref{eq:freeSFRL14} provides an updated calibration to derived SFRs from rest-frame 1.4\,GHz luminosities using a more robust SED fitting technique than those given in \citet{DaviesGalaxyMassAssembly2017}. 

\begin{figure*}
    \centering
    \includegraphics[width = \linewidth]{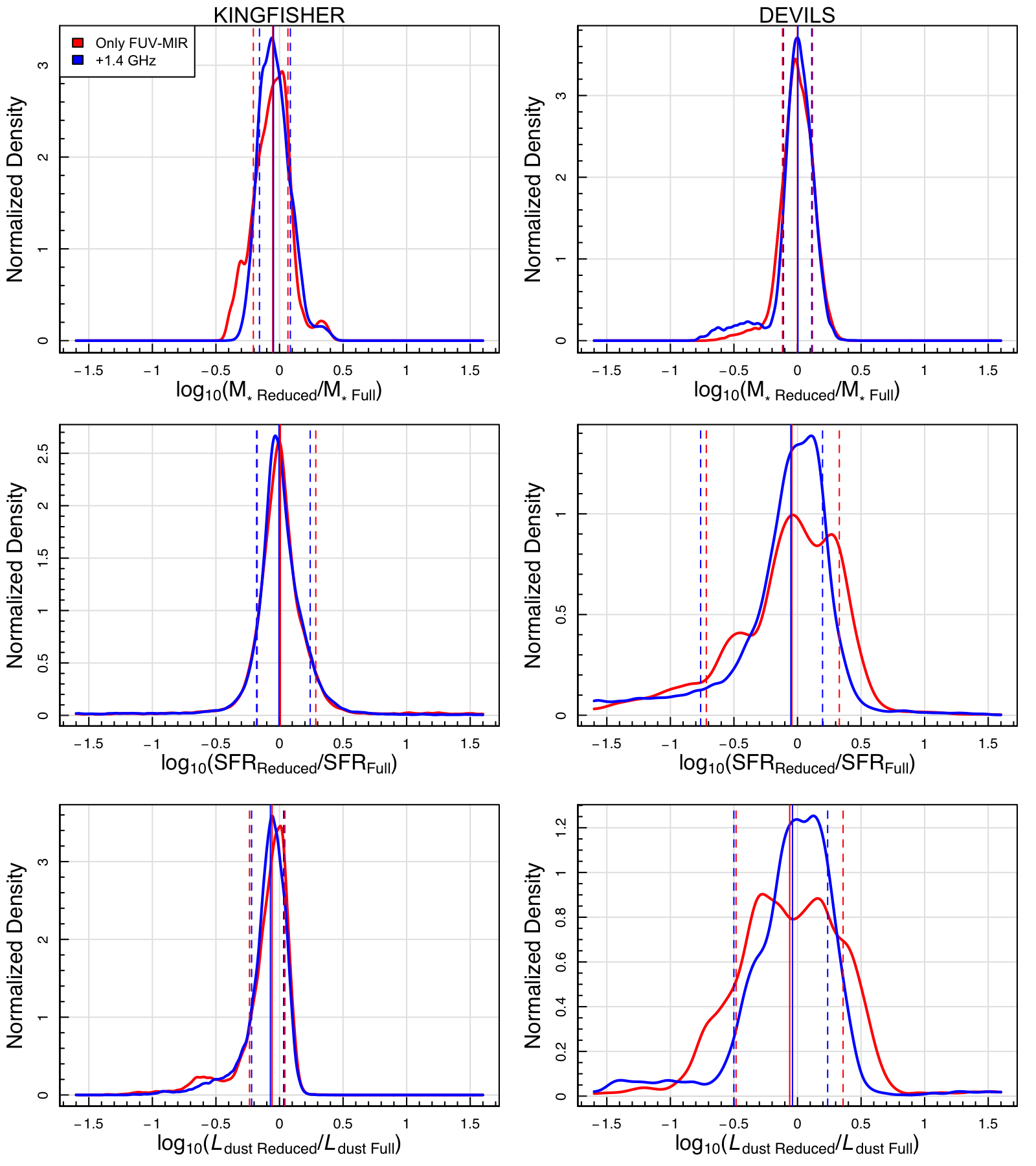}
    \caption[Comparison of the stellar mass, SFR, and dust luminosity for the FUV-MIR and FUV-MIR + 1.4\,GHZ iterations relative to the values derived using the full FUV-radio photometry coverage.]{
    \new{Difference between probability distributions for the stellar mass (top), SFR (middle), and dust luminosity (bottom) derived using the FUV-MIR (red) and FUV-MIR+1.4\,GHz (blue) iterations relative to the probability distributions derived using the full FUV-radio photometry coverage. }
    Due to differences in photometry coverage we show the KINGFISHER (left) and DEVILS (right) samples separately. 
    In each panel the solid line shows the median offset while the 16th and 84th percentiles are shown as the dashed line.  }
    \label{fig:DiffWaveCoverageFUVNIR}
\end{figure*}

\begin{figure*}
    \centering
    \includegraphics[width = \linewidth]{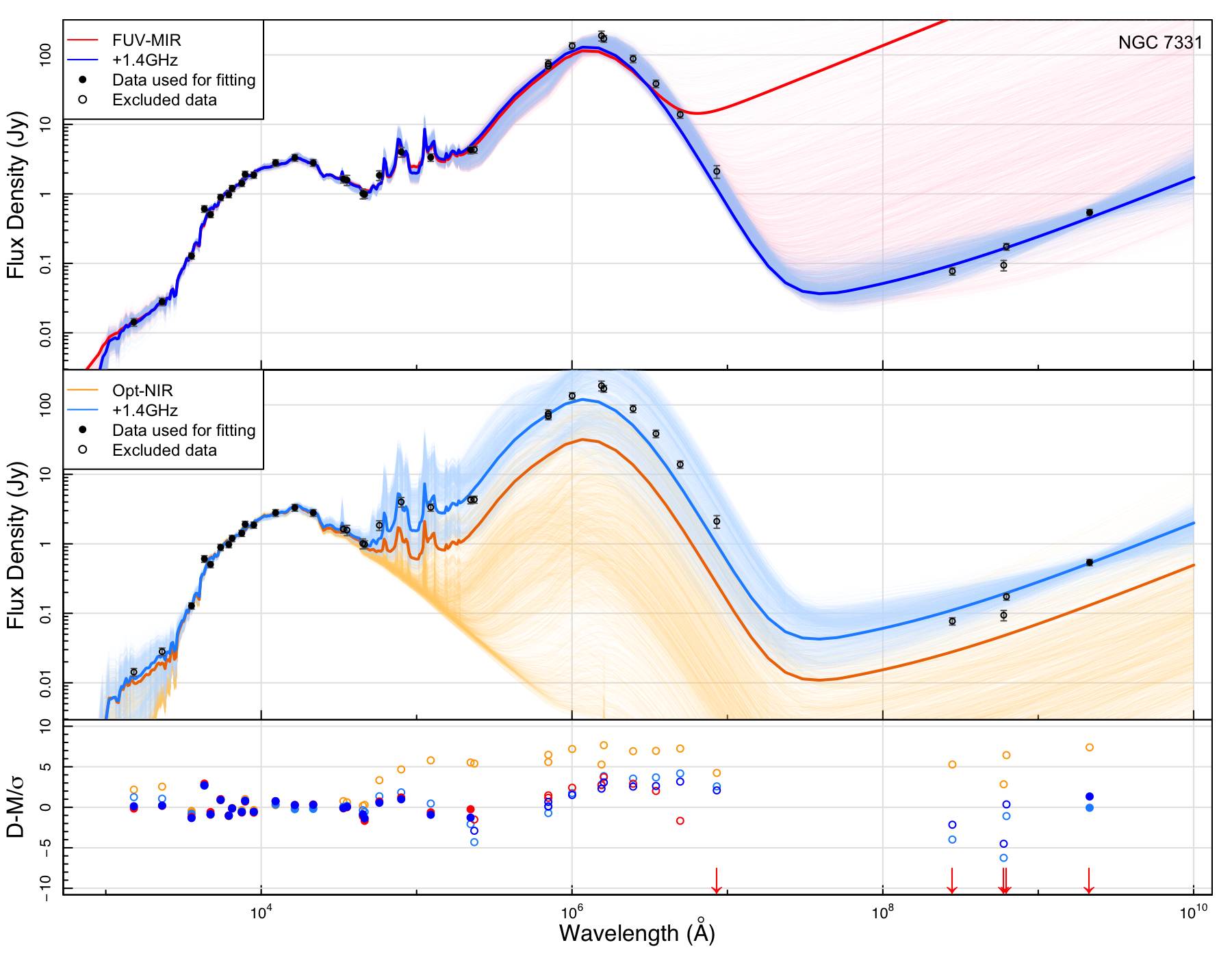}
    \caption[The results of fitting the photometry of NGC 7331 using different wavelength sampling. ]
    {\new{Comparisons of the SED fitting results for NGC 7331 when only FUV-MIR (top, red) or optical-NIR (middle, orange) is used versus when a 1.4\,GHz measurement is also included (blue)
    In each panel, the dark line shows the best fitting (maximum likelihood) SED, while the lighter coloured lines show the sampling of the final MCMC chain.}
    The upper and middle panels show the input photometry for the fits (solid circles) and the photometry that was excluded from the fit for testing purposes (open circles).
    The lower panel shows the relative offset (data - model / error) of each data point from the best-fitting SEDs derived using the different input data. }
    \label{fig:OptNIRCompSED}
\end{figure*}

\begin{figure*}
    \centering
    \includegraphics[width = \linewidth]{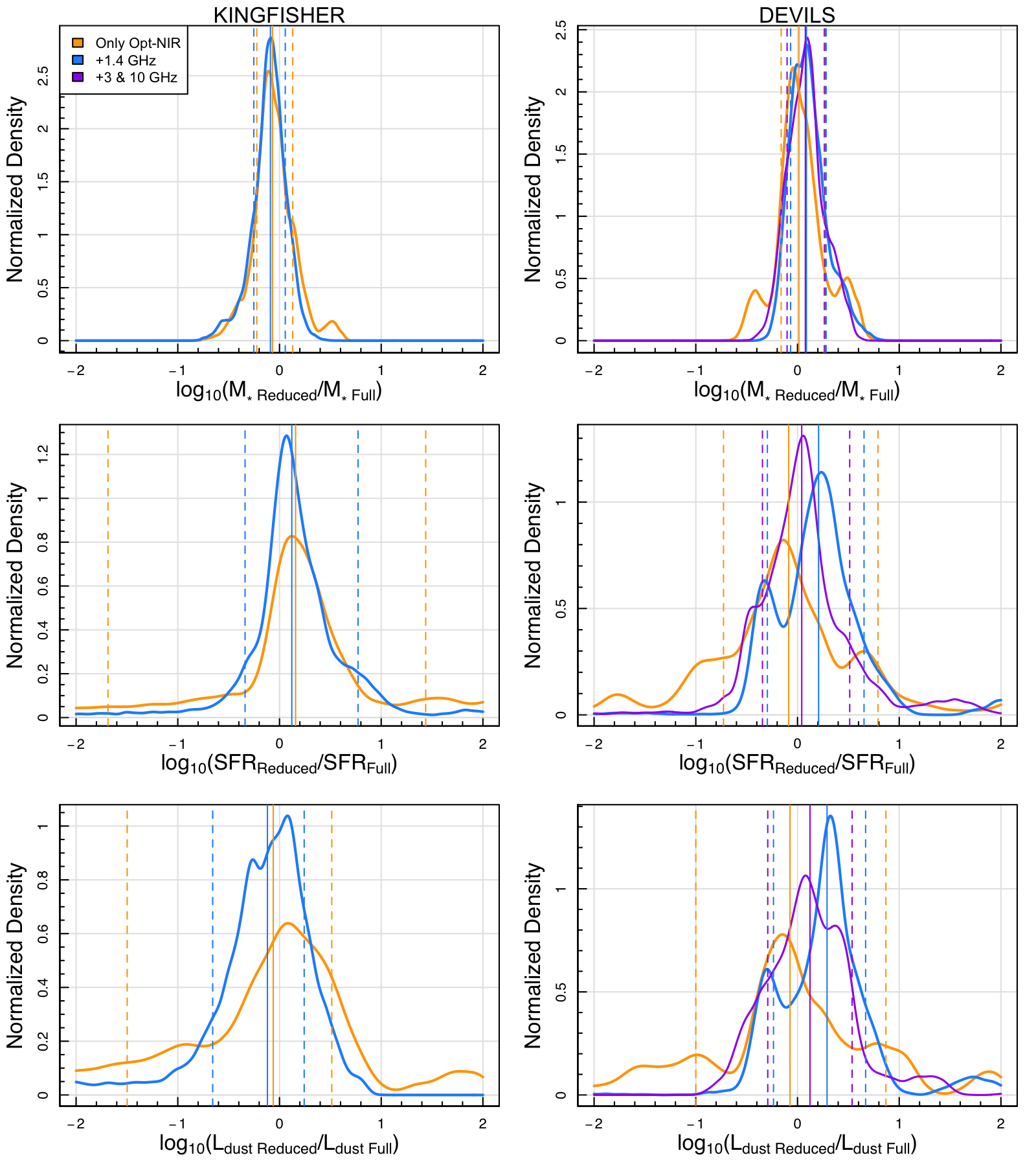}
    \caption[Comparison of the stellar mass, SFR, and dust luminosity for the optical-NIR and optical-NIR + 1.4\,GHZ iterations relative to the values derived using the full FUV-radio photometry coverage.]
    {\new{Difference between probability distributions for the stellar mass (top), SFR (middle), and dust luminosity (bottom) derived using the optical-NIR (orange), optical-NIR+1.4\,GHz (light blue), and optical-NIR+1.4,3,10\,GHz (purple - DEVILS only) iterations relative to the probability distribution derived using the full FUV-radio photometry coverage.}
    Due to differences in photometry coverage we show the KINGFISHER (left) and DEVILS (right) samples separately. 
    In each panel, the solid line shows the median offset while the 16th and 84th percentiles are shown as the dashed line.  }
    \label{fig:DiffWaveCoverageOptMIR}
\end{figure*}

\section{Using the radio continuum as an anchor for energy balance}\label{sec:energybalance}
As we move into the era of the SKA and its precursors, the number of galaxies with radio data will significantly increase \new{(e.g. the Evolutionary Map of the Universe; EMU, \citealt{NorrisEMUEvolutionaryMap2011}, the Widefield ASKAP L-band Legacy All-sky Blind surveY; WALLABY, \citealt{KoribalskiWALLABYSKAPathfinder2020}, the Rapid ASKAP Continuum Survey; RACS-mid, \citealt{McConnellRapidASKAPContinuum2020}, and Deep Investigation of Neutral Gas Origins; DINGO, \citealt{MeyerExploringHIUniverse2009})}. 
As there are no currently operating or planned future far-infrared observatories, there is no way to obtain FIR imaging of galaxies beyond what already exists, and as such techniques that rely on energy balance will be limited to higher redshifts (where the rest-frame FIR is pushed to millimeter wavelengths) or existing samples. 
However, due to the tight infrared--radio correlation, it is theoretically possible to use radio continuum emission to constrain the re-emission of processed stellar light by dust instead of the FIR.
As such, here we explore the potential of using the radio continuum for energy balance when the FIR is unavailable. 
Using the DEVILS and KINGFISHER sample of galaxies we explore the utility of including the radio continuum as an anchor for energy balance in the absence of FIR data. 

As discussed in \citet{ThorneDeepExtragalacticVIsible2022}, FIR photometry is required when SED fitting to constrain a MIR AGN component. 
As we are removing FIR photometry in this section we limit our DEVILS sample to only galaxies with $f_\text{AGN} < 0.1$ from our \textsc{ProSpect} fits and exclude the AGN component from our \textsc{ProSpect} fits (i.e set \texttt{AGNlum}$=0$).

To help constrain the fit to the radio data we implement a prior on the synchrotron spectral index of: 
\begin{equation}\label{eq:asyprior}
    \exp{\left(-\frac{1}{2} \left(\frac{\alpha_\text{sy} - 0.9}{0.2} \right)^2 \right) },
\end{equation}
\neww{where the values for the mean and standard deviation were informed by the distribution of values obtained for the DEVILS and KINGFISHER samples in Section~\ref{sec:RadioSEDParameters}.}
We also implement a prior on the free-free fraction as given by:
\begin{equation}\label{eq:fffprior}
    \exp{ \left(-\frac{1}{2} \left(\frac{\log_{10}(f_\text{ff}) + \log_{10} (0.08)}{0.3} \right)^2 \right) }.
\end{equation}
This corresponds to a mean free-free fraction of 0.08 and a $1\sigma$ range spanning $\sim 0.05-0.20$ as suggested by \citep{CondonRadioemissionnormal1992}. 

Using this implementation we explore two scenarios. 
In the first, we remove all photometry between 24-850 micron (i.e. MIPS 24 - HFI/SCUBA 850\,$\mu$m bands) and limit our radio coverage to just a single measurement at 1.4\,GHz. 
We choose 1.4\,GHz as this is the frequency of a number of SKA precursor surveys such as the MIGHTEE, WALLABY, RACS-mid, and DINGO surveys and as such will provide 1.4\,GHz continuum measurements covering a very large area. 
In this scenario, both samples have good coverage from the FUV-MIR, especially the KINGFISHER sample, allowing for reasonable constraint on balance between attenuated and re-emitted energy.
We also explore a more extreme scenario where only optical and NIR photometry is available. 
In practice, this is implemented by removing measurements at wavelengths longer than IRAC 1 and also removing GALEX UV measurements. 
This is closer to what might be expected outside panchromatic survey fields, but where estimates for the properties of the host galaxies of transient events are still required \citep[e.g.][]{SeymourPKS2250351giant2020,NorrisMeerKATuncoversphysics2022}.

\subsection{Where FUV-MIR data are available}\label{sec:RadioFUVMIR}

To test the utility in including a radio continuum measurement where FIR data are not available we re-fit our galaxy samples removing all FIR photometry between 24-850\,$\mu$m (i.e. MIPS 24 - HFI/SCUBA 850\,$\mu$m bands) and using only the available measurements at 1.4\,GHz.
For a control sample, we also re-fit our sample using only the available FUV-MIR data with no FIR or radio data included. 
To evaluate the improvement when including the 1.4\,GHz measurement we compare to the galaxy properties as derived using the full FUV-radio fits as described in Section~\ref{sec:RadioSED}.

Figure~\ref{fig:DiffWaveCoverageFUVNIR} presents comparisons of the probability distributions for the derived stellar masses, SFRs, and dust luminosities when fitting with different wavelength sampling. 
The red shows the case where no FIR or radio data is being used to constrain the dust emission element of the energy balance, while the blue shows the case where we have included a 1.4\,GHz measurement in the fit. 
The $x$-axis in these figures is the offset from the parameter derived using the fitting process applied using the full available FUV-radio coverage.
\new{
To calculate these differences, for each galaxy and parameter, we randomly sample the posterior 1000 times for both the full and reduced photometry iterations.
This highlights not only the impact on the single best-fitting value but also the distribution of sampled values and therefore the uncertainty on each parameter.
}
We find that for the KINGFISHER sample, the addition of the 1.4\,GHz measurement makes little difference in the recovery of the SFRs and dust luminosities, but does allow for a slightly tighter recovery of stellar masses.
However, for the DEVILS sample, the inclusion of a 1.4\,GHz measurement results in a median offset closer to zero for both the SFR and dust luminosity and results in a tighter distribution. 

The differences in impact between the two surveys can be explained by the excellent wavelength coverage in the MIR for the KINGFISHER sample. 
By combining four WISE channels with four IRAC channels, the SED fits for the KINGFISHER sample are well constrained in the MIR allowing for good constraint on the dust luminosity and SFR. 
However, as there is less MIR wavelength coverage for DEVILS and the photometry has lower SNR, the inclusion of a 1.4\,GHz measurement allows for better recovery of the SFRs and dust luminosities. 

These results suggest that in the absence of FIR photometry, the inclusion of a radio continuum measurement with well-chosen priors can aid in constraining the emission from dust and therefore lead to more accurate estimates of SFR and dust luminosity where FIR data is limited or of low SNR. 
However, the offsets in this scenario are quite moderate. 

\subsection{Where only optical-NIR data are available}
We also explore a more extreme scenario where only optical-NIR photometry is available and a MIR constraint on dust emission is not possible. 
We implement this by removing the \textit{GALEX} UV measurements and all photometry between 3.6 - 850\,$\mu$m (IRAC 1 - SCUBA/HFI 850\,$\mu$m). 
As per Section~\ref{sec:RadioFUVMIR}, we also fit using just the optical-NIR data as a control sample for comparison.
\new{Figure~\ref{fig:OptNIRCompSED} shows the best-fitting SEDs derived for NGC 7331 when only FUV-MIR (upper panel) and optical-NIR (middle panel) photometry is used versus when a 1.4\,GHz measurement is also incorporated. }
We find that when including the 1.4\,GHz radio measurement, the derived SED is in closer agreement with the excluded data, especially in the MIR and FIR.
Though we only show one example here, we find that generally the addition of the 1.4\,GHz measurement results in a best-fitting SED that more closely resembles the actual SED of the galaxy. 
However, in some galaxies, the difference between SEDs is not as large as in the example shown here. 

Figure~\ref{fig:DiffWaveCoverageOptMIR} presents comparisons of the derived stellar masses, SFRs, and dust luminosities when fitting with only the optical-NIR versus fitting with optical-NIR and 1.4\,GHz measurement and is calculated as per Figure~\ref{fig:DiffWaveCoverageFUVNIR}.
In this scenario, we find that the addition of a 1.4\,GHz measurement results in tighter distributions for the SFR and dust luminosity for both DEVILS and KINGFISHER. 
However, it also increases the median offset for the dust luminosities, and for the DEVILS SFR. 
However, these changes in offset are small and within the typical errors for each parameter when fitting with better wavelength coverage. 

We also find that the inclusion of a flux density measurement at 1.4\,GHz decreases the median uncertainty on derived SFRs by 0.1\,dex for both DEVILS and KINGFISHER (from 0.55 to 0.45\,dex for KINGFISHER and from 0.35 to 0.25\,dex for DEVILS where in both cases the median uncertainty when including all FUV-radio data is $\sim0.2$\,dex). 
Similarly, the inclusion of a 1.4\,GHz measurement decreases the median uncertainty on derived dust luminosities 0.2\,dex 
\new{(from 0.84 to 0.64)} for KINGFISHER and 0.14\,dex for DEVILS \new{(from 0.41 to 0.27\,dex)}.

\begin{figure}
    \centering
    \includegraphics[width = \linewidth]{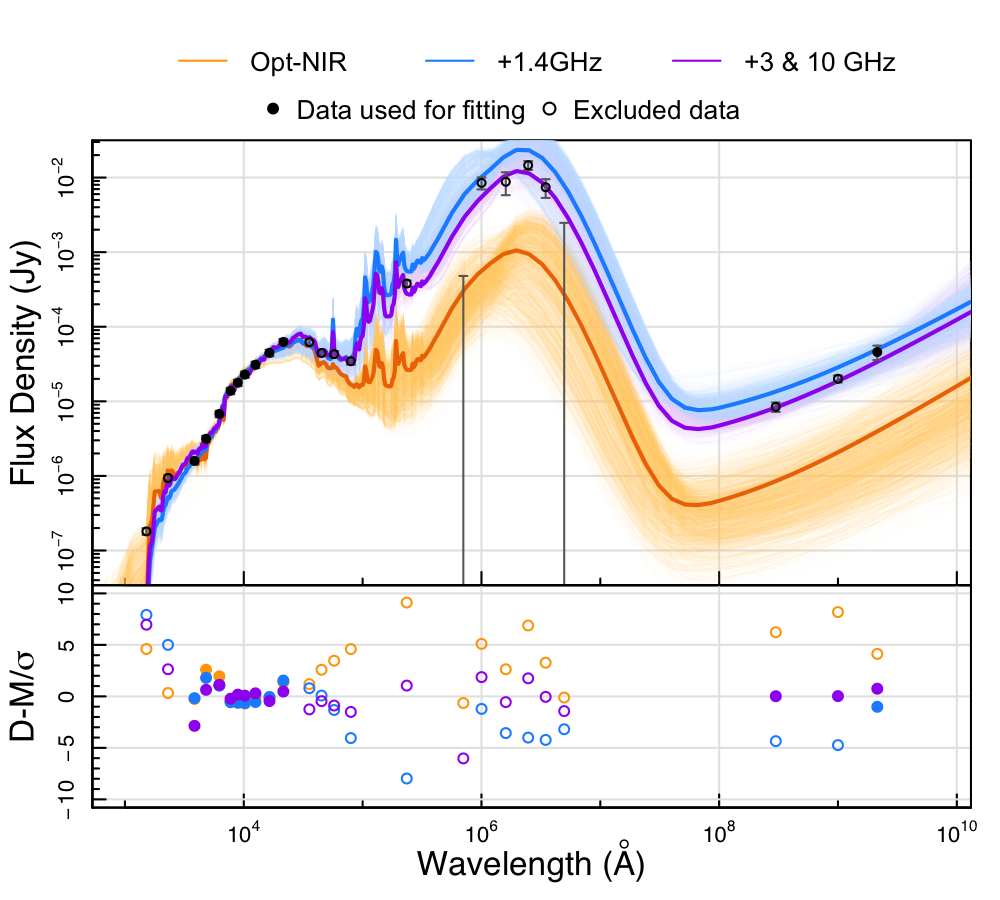}
    \caption{\new{Comparisons of the SED fitting results for an example DEVILS galaxy (101500953042568112) when only the optical-NIR (orange) is used versus when a 1.4\,GHz measurement (blue) or measurements at 1.4, 3, and 10\,GHz are included (purple).  
    The upper panel shows the input photometry for the fits (solid circles) and the photometry that was excluded from the fit for testing purposes (open circles). 
    The lower panel shows the relative offset (data - model / error) of each data point from the best-fitting SEDs derived using the different input data. }}
    \label{fig:DEVILSaddradio}
\end{figure}

\subsubsection{Incorporating additional radio data}
\new{
We also explore the constraint provided by including additional radio data in addition to a 1.4\,GHz measurement. 
To do this we elect to use only the DEVILS sample as the frequency coverage for all galaxies is uniform (i.e. measurements at 1.4, 3, and 10\,GHz for all galaxies).
We explore the situation where only optical and NIR data are available and re-fit the SEDs of the DEVILS galaxies incorporating all three radio measurements. 
As the increased coverage at radio frequencies allows for additional constraint on the two radio parameters (\asy\ and \fff), we adapt the imposed priors as described in Equations~\ref{eq:asyprior} and \ref{eq:fffprior} to be less restrictive by increasing the standard deviations from 0.2 and 0.3 to 0.4 and 0.5 for the \asy\ and \fff\ parameters respectively. 
}

\new{
Figure~\ref{fig:DEVILSaddradio} shows the impact for an example DEVILS galaxy where the fit obtained using just the optical-NIR data is shown in orange, the fit using an additional 1.4\,GHz measurement is shown in blue, and the addition of all three radio measurements is shown in purple.
It is clear that the inclusion of even just the 1.4\,GHz measurement provides substantial improvement in recovering a best-fitting SED consistent with the observed FIR data, however, the incorporation of the additional measurements at 3 and 10\,GHz provides the best match to the observed MIR and FIR. 
We show the impact of including additional radio measurements on the derived galaxy properties as the purple line in Figure~\ref{fig:DiffWaveCoverageOptMIR}. 
While the inclusion of the 3 and 10\,GHz measurements makes little difference to the derivation of stellar masses, it does result in a tighter distribution for the SFRs and dust luminosities despite the less restrictive prior. 
}

Along with the results from the previous Section, these comparisons highlight the utility of including at least one radio continuum measurement to improve the accuracy of derived galaxy SEDs and properties. 
This will be especially important in the era of the SKA where large all-sky surveys will target regions with sparse optical-NIR photometric coverage but will still require accurate estimates of galaxy properties such as SFR and dust luminosities. 

\section{Summary} \label{sec:radiosummary}
In this work, we have extended the \textsc{ProSpect} SED-fitting code to simultaneously model the FUV-FIR regimes and the radio-continuum, incorporating both the free-free and synchrotron components. 
We have applied this extended version of \textsc{ProSpect} to a sample of 30 galaxies from the DEVILS sample and 34 KINGFISHER galaxies to explore the impact of modeling the full FUV-radio SEDs of galaxies. 

We derive free-free fractions (\fff) and synchrotron spectral indices (\asy) for all of our galaxies and find that they are in reasonable agreement with estimates derived when fitting only at radio frequencies from \citet{TabatabaeiRadioSpectralEnergy2017}. 
We recover a median free-free fraction at 1.4\,GHz of $0.05\pm0.03$ and $0.09 \pm 0.04$ and a median synchrotron \new{spectral index} of $-0.84 \pm  0.29$ and $-0.98 \pm 0.22$ for the DEVILS and KINGFISHER samples respectively.
We also find that objects with very low free-free fractions, suggesting higher synchrotron contributions, are also identified as radio-loud AGN using the infrared-radio correlation \citep{WhittamMIGHTEEnatureradioloud2022}. 
These objects are also often associated with higher contributions from an AGN component in the mid-infrared.

\new{We explore the relationship between SFR and rest-frame 1.4\,GHz luminosity and find that most of our sample follows a tight relation, where a higher 1.4\,GHz luminosity is associated with a higher SFR. 
This is in agreement with previous results from \citet{CondonRadioemissionnormal1992,BellEstimatingStarFormation2003,BrownCalibrationultravioletmidinfrared2017} and \cite{DaviesGalaxyMassAssembly2017} at low-$z$.
We also find that the relationship is consistent between the DEVILS and KINGFISHER samples with no obvious discontinuity or change in slope, suggesting no evolution in the relationship for $z<1.5$. 
We recover a steeper slope than \citet{DaviesGalaxyMassAssembly2017} of $m=0.925 \pm 0.08$ consistent with a linear relationship between SFR and 1.4\, luminosity.  
This suggests that emission from both the free-free and synchrotron components scales linearly with star formation. }

Additionally, as the radio continuum extension to \textsc{ProSpect} is parameterised using the relationship between far-infrared and free-free emission, we find that the radio continuum can be a reasonable substitute for FIR photometry for the purposes of energy balance. 
We find that the incorporation of radio continuum data is especially useful in improving the accuracy of derived SFRs and dust luminosities in situations where only optical-NIR photometry would be available otherwise.
This finding is valuable due to the lack of operating and planned far-infrared observatories and to the upcoming plethora of data from the SKA and its precursors (ASKAP and MeerKAT, \citealt{MeyerExploringHIUniverse2009,NorrisEMUEvolutionaryMap2011,JarvisMeerKATInternationalGHz2016,KoribalskiWALLABYSKAPathfinder2020,McConnellRapidASKAPContinuum2020} ).  
\new{We also demonstrate that including additional radio measurements (in this case at 3 and 10\,GHz) allows for additional constraint on the energy balance and therefore the SFR and dust luminosity. }
We recommend that, in situations where FIR data is not available, the radio continuum be used to aid in constraining the energy balance and therefore SFRs and dust luminosities.
This can be implemented either using a single radio measurement with informative priors for the radio parameters or using measurements at multiple radio frequencies with less restrictive or uniform priors.

In addition to the benefits outlined above, extending FUV-FIR SED fitting techniques to radio frequencies, even in a predictive mode, could be used to improve radio source-finding techniques and could aid in associating radio sources with optical counterparts. 
We have demonstrated that for normal galaxies, the derived free-free fractions are relatively constant with redshift and other galaxy properties. 
Hence, using default values ($f_\text{ff} = 0.1$ and $\alpha_\text{sy} = -0.8$), \textsc{ProSpect} could be used to predict the radio emission from a given SED fit which could be used alongside positional information to correctly associate radio sources with their optical counterpart. 
Additionally, if sources lacking FIR photometry have significantly more radio emission than predicted by \textsc{ProSpect}, this could be used to identify radio-loud AGN in situations where the infrared--radio correlation cannot be used directly. 
These scenarios all highlight the potential advantages of simultaneous and self-consistent consideration of the FUV-radio SEDs of galaxies.

Although this work considers only radio frequencies above 1.3\,GHz, the radio frequency coverage in \textsc{ProSpect} could be extended to model radio emission at lower frequencies (i.e. $\sim150\,$MHz from surveys such as the Low Frequency Array (LOFAR) Two Metre Sky Survey; LOTSS \citealt{SmithLOFARTwometreSky2021,ShimwellLOFARTwometreSky2022,HeesenNearbygalaxiesLOFAR2022}).
However, to do this, \textsc{ProSpect} may need to be adapted to account for potential flattening of the radio continuum below $\sim1\,$GHz \citep{CondonRadioemissionnormal1992, DelhaizeVLACOSMOSGHzlarge2017, Galvinspectralenergydistribution2018, DeyLowfrequencyRadioContinuum2022}.

\section*{Acknowledgements}
We thank the anonymous referee for their constructive report. 
JET is supported by the Australian Government Research Training Program (RTP) Scholarship.
ASGR and LJMD acknowledge support from the \textit{Australian Research Council's} Future Fellowship scheme (FT200100375 and FT200100055 respectively). 
SB acknowledges support from the \textit{Australian Research Council’s} Discovery Project and Future Fellowship funding schemes (DP180103740, FT200100375).

We acknowledge the traditional owners of the land on which this research was completed, the Whadjuk Noongar people, and the land on which the AAT stands, the Gamilaraay people, and pay our respects to elders past and present.

DEVILS is an Australian project based around a spectroscopic campaign using the Anglo-Australian Telescope. 
The DEVILS input catalogue is generated from data taken as part of the ESO VISTA-VIDEO \citep{JarvisVISTADeepExtragalactic2013} and UltraVISTA \citep{McCrackenUltraVISTAnewultradeep2012} surveys. DEVILS is part funded via Discovery Programs by the Australian Research Council and the participating institutions. The DEVILS website is \url{https://devilsurvey.org}. The DEVILS data is hosted and provided by AAO Data Central (\url{https://datacentral.org.au/}).

This work was supported by resources provided by the Pawsey Supercomputing Centre with funding from the Australian Government and the Government of Western Australia. 

All of the work presented here was made possible by the free and open R software environment \citep{RCoreTeamLanguageEnvironmentStatistical2020}. All figures in this paper were made using the R \textsc{magicaxis} package \citep{RobothammagicaxisPrettyscientific2016}. This work also makes use of the \textsc{celestial} package \citep{RobothamCelestialCommonastronomical2016}.

\section*{Data Availability}
The DEVILS data products used in this paper are presented in \cite{DaviesDeepExtragalacticVIsible2021}, \cite{ThorneDeepExtragalacticVIsible2021} and \cite{ThorneDeepExtragalacticVIsible2022}. 
They will be made publicly available as part of the DEVILS first data release described in Davies et al. (in preparation).
The galaxy properties derived in this work  for the KINGFISHER sample are available in Appendix~\ref{app:radioSEDs} and made available as supplementary material. 
We also make available individual figures for each KINGFISHER galaxy showing the best fitting SED and the resulting star formation and metallicity history. 
Any other outputs for the KINGFISHER sample will be made available upon reasonable request to the corresponding author. 




\bibliographystyle{mnras}
\bibliography{MyBib} 



\appendix

\newcommand{\grasil}{\textsc{grasil}}
\newcommand{\magphys}{\textsc{magphys}}
\newcommand{\cigale}{\textsc{cigale}}

\section{Fits to the KINGFISHER galaxy sample}\label{app:radioSEDs}

\begin{figure*}
    \centering
    \includegraphics[width = 0.8\linewidth]{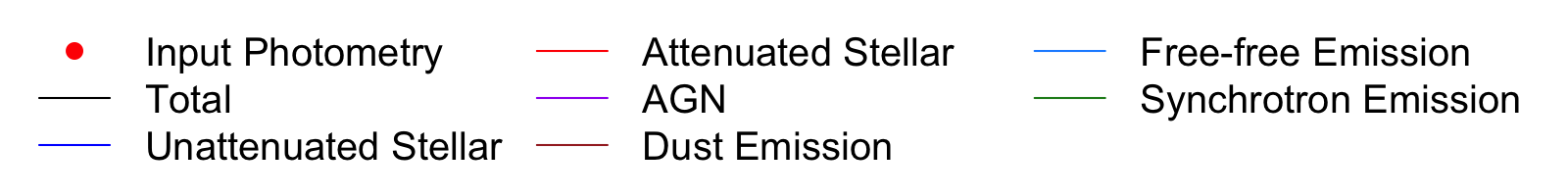} \\
    \includegraphics[width = 0.8\linewidth]{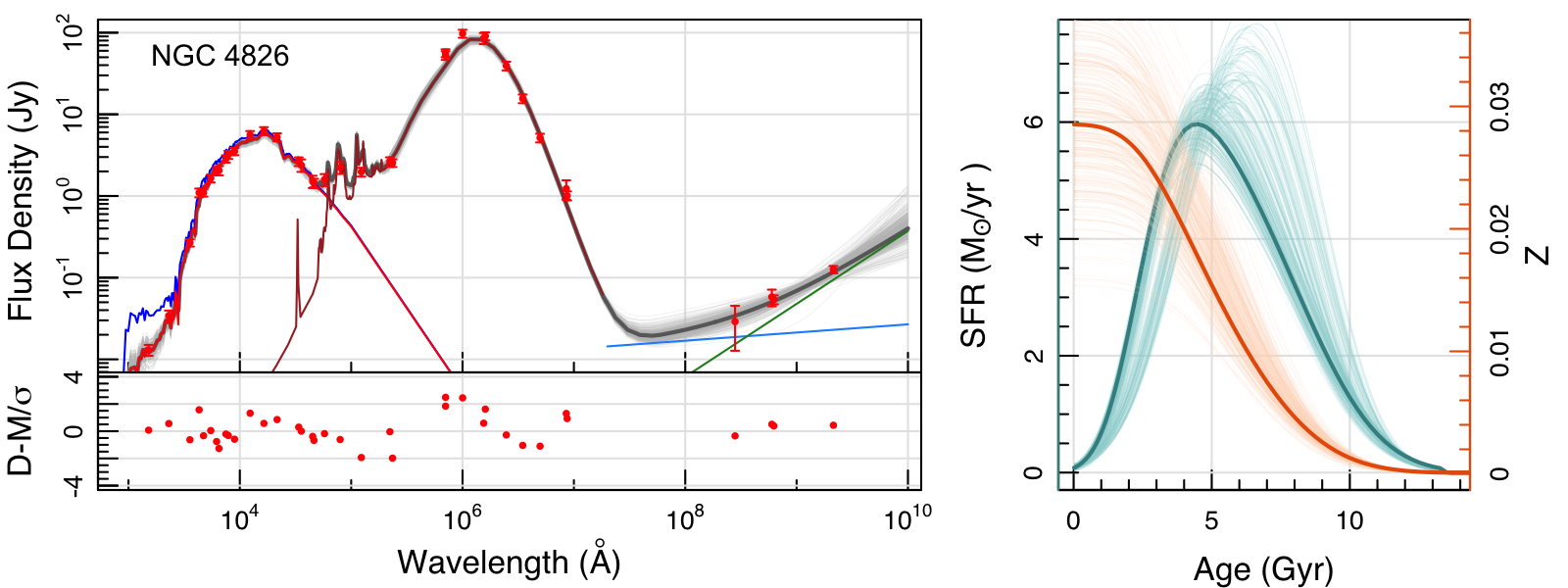}
    \caption{Example of FUV-radio SED fit to NGC 4826 from the KINGFISHER sample based on the measurements from \citet{DaleUpdated34bandPhotometry2017} and \citet{TabatabaeiRadioSpectralEnergy2017} (red) overlaid with the best fitting \textsc{ProSpect} total SED (black), stellar unobscured (blue) and obscured (red), dust (brown), radio synchrotron (green) and free-free (light blue) emission. 
    The sampling of the final MCMC posterior is shown in (grey) to illustrate the uncertainty in the fits. 
    The lower left panel shows the residuals for each measurement (data - model / error). 
    The right panel shows the best-fit SFH (green) and metallicity history (orange) along with the sampling of the posterior for both histories (lighter shades).
    Equivalent figures for the rest of the sample are made available as supplementary material. }
    \label{fig:ExampleKFSED}
\end{figure*}

This Appendix presents the \textsc{ProSpect} fits to the sample of 24 KINGFISHER galaxies as described in Section~\ref{sec:RadioSED}. 
The physical quantities for each of the galaxies are given in Table~\ref{tab:KINGFISHERResults} and provided as supplementary material. 
Figure~\ref{fig:ExampleKFSED} shows the resulting SED fit, star formation and metallicity history for an example galaxy (NGC 4826) where equivalent figures for the rest of the KINGFISHER sample are available as supplementary material. 

\begin{table*}
    \centering
    \caption[Parameters derived by \textsc{ProSpect} for the sample of KINGFISHER galaxies]{Parameters derived by \textsc{ProSpect} for the sample of KINGFISHER galaxies. Full table available online.}
    \label{tab:KINGFISHERResults}
    \begin{tabular}{l c c c c c c }
\hline

Galaxy	&	$\log_{10} \left( \frac{M_\star}{M_\odot} \right) $ 				&	$\log_{10} 
 \left( \frac{\text{SFR}}{M_\odot \text{yr}^{-1}} \right)	$ &	 $\log_{10} \left(\frac{M_\text{dust}}{M_\odot}\right) $&	$\log_{10} \left(\frac{L_\text{dust}}{L_\odot}\right)$					&	$\log_{10} \left(f_\text{ff}\right)$				&	$\alpha_\text{sy}$			\\	[6pt] 
\hline 
IC2574	&$	8.53	_{-	0.25	}^{+	0.09	}$ & $	-1.17	_{-	0.12	}^{+	0.08	}$ & $	5.97	_{-	0.18	}^{+	0.18	}$ & $	8.29	_{-	0.08	}^{+	0.07	}$ & $	-0.83	_{-	0.43	}^{+	0.33	}$ & $	-0.49	_{-	1.09	}^{+	0.49	}$\\	[3pt]
NGC 0337	&$	9.60	_{-	0.07	}^{+	0.12	}$ & $	0.02	_{-	0.30	}^{+	0.09	}$ & $	7.56	_{-	0.14	}^{+	0.15	}$ & $	10.01	_{-	0.09	}^{+	0.07	}$ & $	-1.28	_{-	0.20	}^{+	0.35	}$ & $	-1.30	_{-	0.36	}^{+	0.54	}$\\	[3pt]
NGC 0628	&$	10.01	_{-	0.13	}^{+	0.09	}$ & $	-0.24	_{-	0.11	}^{+	0.21	}$ & $	7.57	_{-	0.04	}^{+	0.27	}$ & $	9.86	_{-	0.08	}^{+	0.05	}$ & $	-0.88	_{-	0.14	}^{+	0.27	}$ & $	-1.02	_{-	0.43	}^{+	0.52	}$\\	[3pt]
NGC 1266	&$	10.59	_{-	0.08	}^{+	0.08	}$ & $	-1.30	_{-	20.02	}^{+	0.48	}$ & $	7.29	_{-	0.87	}^{+	0.14	}$ & $	10.23	_{-	0.20	}^{+	0.08	}$ & $	-1.39	_{-	0.30	}^{+	0.26	}$ & $	-1.05	_{-	0.36	}^{+	0.40	}$\\	[3pt]
NGC 1482	&$	10.49	_{-	0.09	}^{+	0.07	}$ & $	-0.17	_{-	0.76	}^{+	0.28	}$ & $	7.66	_{-	0.11	}^{+	0.19	}$ & $	10.63	_{-	0.06	}^{+	0.06	}$ & $	-1.08	_{-	0.26	}^{+	0.23	}$ & $	-1.20	_{-	0.38	}^{+	0.39	}$\\	[3pt]
NGC 2146	&$	10.96	_{-	0.13	}^{+	0.07	}$ & $	0.56	_{-	0.05	}^{+	0.35	}$ & $	7.92	_{-	0.18	}^{+	0.31	}$ & $	11.04	_{-	0.08	}^{+	0.11	}$ & $	-1.11	_{-	0.12	}^{+	0.18	}$ & $	-0.87	_{-	0.25	}^{+	0.17	}$\\	[3pt]
NGC 2798	&$	10.33	_{-	0.13	}^{+	0.04	}$ & $	0.29	_{-	0.09	}^{+	0.12	}$ & $	7.39	_{-	0.15	}^{+	0.20	}$ & $	10.54	_{-	0.09	}^{+	0.03	}$ & $	-0.75	_{-	0.37	}^{+	0.19	}$ & $	-0.82	_{-	0.60	}^{+	0.48	}$\\	[3pt]
NGC 2841	&$	11.11	_{-	0.05	}^{+	0.06	}$ & $	-26.10	_{-	55.75	}^{+	20.02	}$ & $	8.05	_{-	0.16	}^{+	0.07	}$ & $	10.00	_{-	0.06	}^{+	0.05	}$ & $	-1.11	_{-	0.19	}^{+	0.32	}$ & $	-1.01	_{-	0.74	}^{+	0.83	}$\\	[3pt]
NGC 2976	&$	9.26	_{-	0.04	}^{+	0.12	}$ & $	-1.31	_{-	0.30	}^{+	0.19	}$ & $	6.59	_{-	0.21	}^{+	0.15	}$ & $	8.87	_{-	0.08	}^{+	0.05	}$ & $	-1.04	_{-	0.23	}^{+	0.49	}$ & $	-1.07	_{-	0.83	}^{+	0.68	}$\\	[3pt]
NGC 3049	&$	9.50	_{-	0.14	}^{+	0.04	}$ & $	-0.48	_{-	0.12	}^{+	0.12	}$ & $	7.14	_{-	0.30	}^{+	0.13	}$ & $	9.54	_{-	0.08	}^{+	0.05	}$ & $	-0.79	_{-	0.27	}^{+	0.29	}$ & $	-1.13	_{-	1.07	}^{+	1.13	}$\\	[3pt]
NGC 3077	&$	9.65	_{-	0.21	}^{+	0.04	}$ & $	-23.08	_{-	11.98	}^{+	21.27	}$ & $	6.26	_{-	0.15	}^{+	0.13	}$ & $	8.84	_{-	0.07	}^{+	0.04	}$ & $	-0.50	_{-	0.22	}^{+	0.00	}$ & $	-0.54	_{-	0.85	}^{+	0.54	}$\\	[3pt]
NGC 3184	&$	10.14	_{-	0.09	}^{+	0.06	}$ & $	0.14	_{-	0.07	}^{+	0.06	}$ & $	7.77	_{-	0.12	}^{+	0.19	}$ & $	9.95	_{-	0.07	}^{+	0.04	}$ & $	-0.83	_{-	0.12	}^{+	0.15	}$ & $	-1.00	_{-	1.20	}^{+	0.46	}$\\	[3pt]
NGC 3190	&$	10.86	_{-	0.08	}^{+	0.03	}$ & $	-151.19	_{-	\text{Inf}	}^{+	132.72	}$ & $	7.56	_{-	0.10	}^{+	0.12	}$ & $	9.75	_{-	0.05	}^{+	0.07	}$ & $	-1.13	_{-	0.24	}^{+	0.31	}$ & $	-0.93	_{-	0.71	}^{+	0.57	}$\\	[3pt]
NGC 3265	&$	9.48	_{-	0.09	}^{+	0.03	}$ & $	-0.90	_{-	0.08	}^{+	0.19	}$ & $	6.47	_{-	0.15	}^{+	0.29	}$ & $	9.37	_{-	0.08	}^{+	0.08	}$ & $	-1.05	_{-	0.18	}^{+	0.55	}$ & $	-1.02	_{-	0.52	}^{+	1.02	}$\\	[3pt]
NGC 3627	&$	10.68	_{-	0.07	}^{+	0.09	}$ & $	-0.04	_{-	0.58	}^{+	0.16	}$ & $	7.86	_{-	0.08	}^{+	0.22	}$ & $	10.38	_{-	0.08	}^{+	0.05	}$ & $	-1.09	_{-	0.08	}^{+	0.48	}$ & $	-1.28	_{-	0.48	}^{+	0.99	}$\\	[3pt]
NGC 3938	&$	10.30	_{-	0.07	}^{+	0.10	}$ & $	0.41	_{-	0.15	}^{+	0.05	}$ & $	7.93	_{-	0.10	}^{+	0.14	}$ & $	10.20	_{-	0.05	}^{+	0.04	}$ & $	-0.93	_{-	0.20	}^{+	0.35	}$ & $	-1.08	_{-	1.00	}^{+	0.72	}$\\	[3pt]
NGC 4236	&$	9.01	_{-	0.08	}^{+	0.21	}$ & $	-0.92	_{-	0.56	}^{+	0.13	}$ & $	6.31	_{-	0.30	}^{+	0.29	}$ & $	8.65	_{-	0.12	}^{+	0.07	}$ & $	-1.02	_{-	0.56	}^{+	0.52	}$ & $	-0.98	_{-	0.94	}^{+	0.82	}$\\	[3pt]
NGC 4254	&$	10.42	_{-	0.26	}^{+	0.03	}$ & $	0.53	_{-	0.06	}^{+	0.15	}$ & $	8.17	_{-	0.13	}^{+	0.12	}$ & $	10.56	_{-	0.08	}^{+	0.04	}$ & $	-1.12	_{-	0.17	}^{+	0.14	}$ & $	-0.94	_{-	0.39	}^{+	0.25	}$\\	[3pt]
NGC 4321	&$	10.66	_{-	0.07	}^{+	0.04	}$ & $	0.49	_{-	0.12	}^{+	0.09	}$ & $	8.10	_{-	0.10	}^{+	0.28	}$ & $	10.47	_{-	0.08	}^{+	0.06	}$ & $	-1.03	_{-	0.27	}^{+	0.49	}$ & $	-1.02	_{-	0.50	}^{+	0.88	}$\\	[3pt]
NGC 4536	&$	10.35	_{-	0.17	}^{+	0.07	}$ & $	0.23	_{-	0.09	}^{+	0.15	}$ & $	7.72	_{-	0.13	}^{+	0.29	}$ & $	10.31	_{-	0.08	}^{+	0.04	}$ & $	-0.97	_{-	0.24	}^{+	0.26	}$ & $	-1.01	_{-	0.39	}^{+	0.40	}$\\	[3pt]
NGC 4559	&$	9.49	_{-	0.10	}^{+	0.12	}$ & $	-0.35	_{-	0.05	}^{+	0.12	}$ & $	7.19	_{-	0.16	}^{+	0.14	}$ & $	9.42	_{-	0.06	}^{+	0.06	}$ & $	-1.00	_{-	0.14	}^{+	0.13	}$ & $	-0.98	_{-	1.22	}^{+	0.40	}$\\	[3pt]
NGC 4569	&$	10.41	_{-	0.10	}^{+	0.03	}$ & $	-0.83	_{-	0.22	}^{+	0.39	}$ & $	7.42	_{-	0.11	}^{+	0.21	}$ & $	9.67	_{-	0.07	}^{+	0.05	}$ & $	-1.23	_{-	0.24	}^{+	0.17	}$ & $	-0.83	_{-	1.37	}^{+	0.25	}$\\	[3pt]
NGC 4579	&$	10.95	_{-	0.04	}^{+	0.08	}$ & $	-0.31	_{-	0.63	}^{+	0.10	}$ & $	7.84	_{-	0.15	}^{+	0.12	}$ & $	9.98	_{-	0.06	}^{+	0.06	}$ & $	-1.42	_{-	0.16	}^{+	0.40	}$ & $	-0.46	_{-	0.37	}^{+	0.46	}$\\	[3pt]
NGC 4594	&$	11.13	_{-	0.03	}^{+	0.06	}$ & $	-4.70	_{-	14.62	}^{+	2.94	}$ & $	7.52	_{-	0.12	}^{+	0.10	}$ & $	9.49	_{-	0.06	}^{+	0.06	}$ & $	-1.37	_{-	0.28	}^{+	0.17	}$ & $	0.00	_{-	0.40	}^{+	0.00	}$\\	[3pt]
NGC 4631	&$	9.89	_{-	0.16	}^{+	0.20	}$ & $	0.38	_{-	0.11	}^{+	0.05	}$ & $	7.90	_{-	0.16	}^{+	0.15	}$ & $	10.30	_{-	0.07	}^{+	0.05	}$ & $	-1.18	_{-	0.16	}^{+	0.20	}$ & $	-0.78	_{-	0.25	}^{+	0.33	}$\\	[3pt]
NGC 4725	&$	10.81	_{-	0.15	}^{+	0.03	}$ & $	-0.27	_{-	0.25	}^{+	0.30	}$ & $	7.84	_{-	0.11	}^{+	0.10	}$ & $	9.82	_{-	0.07	}^{+	0.05	}$ & $	-1.13	_{-	0.16	}^{+	0.17	}$ & $	-1.03	_{-	0.39	}^{+	0.29	}$\\	[3pt]
NGC 4736	&$	10.34	_{-	0.02	}^{+	0.07	}$ & $	-0.42	_{-	0.42	}^{+	0.08	}$ & $	7.08	_{-	0.08	}^{+	0.20	}$ & $	9.73	_{-	0.13	}^{+	0.04	}$ & $	-0.73	_{-	0.22	}^{+	0.10	}$ & $	-0.98	_{-	0.59	}^{+	0.53	}$\\	[3pt]
NGC 4826	&$	10.29	_{-	0.06	}^{+	0.10	}$ & $	-1.03	_{-	0.39	}^{+	0.24	}$ & $	7.05	_{-	0.12	}^{+	0.12	}$ & $	9.49	_{-	0.08	}^{+	0.06	}$ & $	-0.71	_{-	0.19	}^{+	0.21	}$ & $	-0.89	_{-	0.95	}^{+	0.49	}$\\	[3pt]
NGC 5055	&$	10.75	_{-	0.09	}^{+	0.03	}$ & $	-0.01	_{-	0.18	}^{+	0.19	}$ & $	7.99	_{-	0.10	}^{+	0.24	}$ & $	10.27	_{-	0.11	}^{+	0.04	}$ & $	-0.90	_{-	0.21	}^{+	0.08	}$ & $	-1.25	_{-	0.33	}^{+	0.63	}$\\	[3pt]
NGC 5457	&$	10.34	_{-	0.10	}^{+	0.04	}$ & $	0.50	_{-	0.10	}^{+	0.07	}$ & $	8.03	_{-	0.15	}^{+	0.16	}$ & $	10.28	_{-	0.06	}^{+	0.06	}$ & $	-0.93	_{-	0.20	}^{+	0.18	}$ & $	-0.73	_{-	0.73	}^{+	0.35	}$\\	[3pt]
NGC 5713	&$	10.46	_{-	0.08	}^{+	0.10	}$ & $	0.26	_{-	0.26	}^{+	0.08	}$ & $	7.74	_{-	0.11	}^{+	0.18	}$ & $	10.48	_{-	0.07	}^{+	0.03	}$ & $	-0.96	_{-	0.24	}^{+	0.15	}$ & $	-0.97	_{-	0.36	}^{+	0.42	}$\\	[3pt]
NGC 5866	&$	10.63	_{-	0.10	}^{+	0.13	}$ & $	-1.90	_{-	7.43	}^{+	0.67	}$ & $	7.35	_{-	0.18	}^{+	0.15	}$ & $	9.43	_{-	0.11	}^{+	0.12	}$ & $	-1.16	_{-	0.66	}^{+	0.54	}$ & $	-0.65	_{-	1.25	}^{+	0.65	}$\\	[3pt]
NGC 6946	&$	10.64	_{-	0.13	}^{+	0.11	}$ & $	0.25	_{-	0.11	}^{+	0.30	}$ & $	8.05	_{-	0.11	}^{+	0.18	}$ & $	10.50	_{-	0.05	}^{+	0.06	}$ & $	-0.96	_{-	0.32	}^{+	0.38	}$ & $	-0.67	_{-	0.56	}^{+	0.60	}$\\	[3pt]
NGC 7331	&$	11.03	_{-	0.15	}^{+	0.01	}$ & $	0.26	_{-	0.16	}^{+	0.23	}$ & $	8.32	_{-	0.11	}^{+	0.16	}$ & $	10.64	_{-	0.07	}^{+	0.05	}$ & $	-1.12	_{-	0.17	}^{+	0.27	}$ & $	-1.32	_{-	0.41	}^{+	0.45	}$\\	[3pt]
\hline
    \end{tabular}
\end{table*}

\section{Comparison of derived KINGFISH galaxy properties}\label{app:KINGFISHERSEDComparisons}

The SEDs of the 61 KINGFISH galaxies were previously fit by \cite{HuntComprehensivecomparisonmodels2019} using the FUV-sub millimeter (850\,$\mu$m) photometry presented in \citet{DaleUpdated34bandPhotometry2017}.
The fitting was performed using \cigale\ \citep{CIGALE,BoquienCIGALEpythonCode2019}, \magphys\  \citep{MAGPHYS}, and \grasil\ \citep{SilvaModelingEffectsDust1998} each adopting the \cite{BruzualStellarpopulationsynthesis2003} SSPs with a \citet{ChabrierGalacticStellarSubstellar2003} IMF. 
For more details about the specific implementations of each model, see \citet{HuntComprehensivecomparisonmodels2019} however we briefly describe each method below.

In the cases of \cigale\ and \magphys, simple parameteric SFHs are used with an exponential plus burst parmeterisation for \magphys\ and a delayed exponential with truncation used for \cigale. 
For \grasil, the SFR at any given time is proportional to the available gas mass where the star formation efficiency and exponential folding timescale of the infalling gas are modelled as free parameters.
When fitting with \cigale, a constant $Z=0.02$ metallicity is adopted while \magphys\ assumes a constant metallicity but the value of which is fit as a free parameter. 
As \grasil\ models the chemical evolution of galaxies, the metallicity of a galaxy within the model evolves over time and is modelled by a free parameter. 
To model the attenuation by dust \magphys\ adopts the two-component model from \citet{CharlotSimpleModelAbsorption2000} while \cigale\ uses a modified star burst attenuation law (e.g. \citealt{CalzettiDustContentOpacity2000}) with differing redenning for stellar populations of different ages (essentially the same assumption as the two-component \citealt{CharlotSimpleModelAbsorption2000} model).
\grasil\ uses gemometry-dependent radiative transfer assuming the \citet{LaorSpectroscopicconstraintsproperties1993} grain opacities to model both the effects of attenuation and emission by dust. 
Re-emission by dust in the FIR is assumed to follow energy-balance in both \magphys\ and \cigale\ however \cigale\ uses the dust models of \citet{DraineDustMassesPAH2007,DraineAndromedaDust2014} while \magphys\ assumes a modified grey-body. 

\begin{figure}
    \centering
    \includegraphics[width = \linewidth]{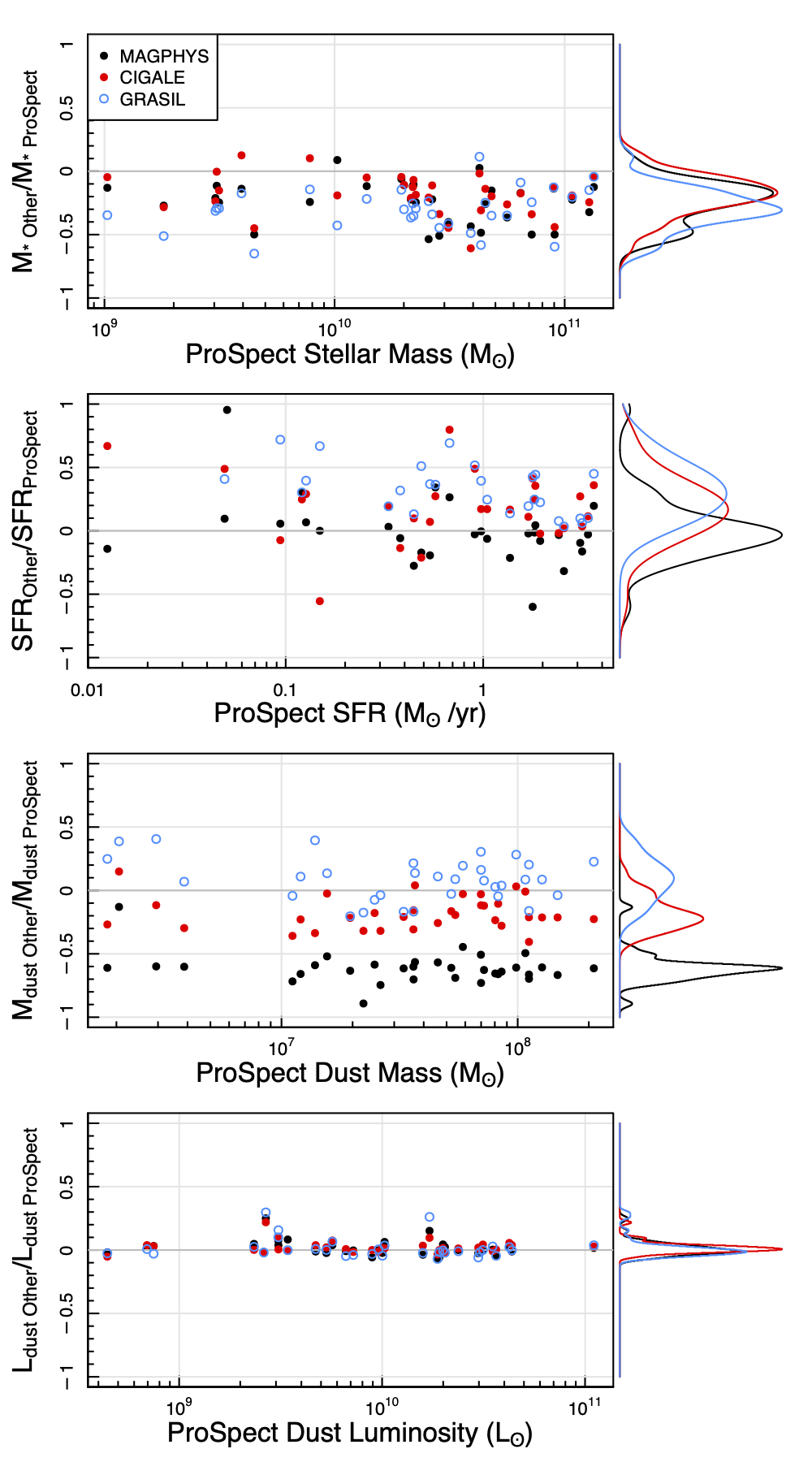}
    \caption[Comparisons of galaxy properties derived using \textsc{ProSpect} for the KINGFISH sample to those presented in \citet{HuntComprehensivecomparisonmodels2019}]{Comparisons of stellar mass, SFR, dust mass, and dust luminosity derived using \textsc{ProSpect} for the KINGFISH sample to those presented in \citet{HuntComprehensivecomparisonmodels2019} using \grasil\ (blue), \cigale\ (red), and \magphys\ (black).
    The right panel shows the projected density for each comparison. }
    \label{fig:KINGFISHSEDComparisons}
\end{figure}

Figure~\ref{fig:KINGFISHSEDComparisons} shows comparisons of the stellar masses, SFRs, dust masses, and dust luminosities as derived by \citet{HuntComprehensivecomparisonmodels2019} to those presented in this work. 
For each parameter, we compare to the properties as derived by \magphys\ (black), \cigale\ (red), and \grasil\ (blue).
We find that \textsc{ProSpect} recovers systematically higher stellar masses when compared to \magphys, \cigale, and \grasil. 
As discussed in \citet{ThorneDeepExtragalacticVIsible2021} this is predominantly the result of a more flexible SFH implementation but also the inclusion of an evolving metallicity history. 
We also find that the use of a more flexible SFH results in systematically lower SFRs, especially when compared to \grasil.
The recovered dust luminosities are in good agreement, however, this does not translate to a good agreement in the dust masses due to different assumptions of the mass-to-light ratio which is highlighted by the clear offsets between SED fitting codes. 
The mass-to-light ratio assumed by \grasil\ results in systematically larger dust masses than \magphys, \cigale, or \textsc{ProSpect}. 

\bsp	
\label{lastpage}
\end{document}